\documentclass[draft,tightenlines,nofootinbib,preprint,aps,eqsecnum,amsmath,amssymb]{revtex4}

\begin{document}

\newcommand{\beq}{\begin{equation}}
\newcommand{\eeq}{\end{equation}}
\newcommand{\bea}{\begin{eqnarray}}
\newcommand{\eea}{\end{eqnarray}}
\newcommand{\cir}{{\buildrel \circ \over =}}

\title{
The Physical Role of Gravitational and Gauge Degrees of Freedom in
General Relativity - II : Dirac versus Bergmann Observables and
the Objectivity of Space-Time}

\medskip

\author{Luca Lusanna}

\affiliation{ Sezione INFN di Firenze\\ Polo Scientifico\\ Via Sansone 1\\
50019 Sesto Fiorentino (FI), Italy\\ E-mail LUSANNA@FI.INFN.IT}

\author{Massimo Pauri}

\affiliation{
Dipartimento di Fisica - Sezione Teorica\\ Universita' di Parma\\
Parco Area Scienze 7/A\\ 43100 Parma, Italy \\E-mail
PAURI@PR.INFN.IT}

\begin{abstract}

This is the second of a couple of papers in which we aim to show
the peculiar capability of the Hamiltonian ADM formulation of
metric gravity to grasp a series of conceptual and technical
problems that appear to have not been directly discussed so far.
In this paper we also propose new viewpoints about issues that,
being deeply rooted into the foundational level of Einstein
theory, seem particularly worth of clarification in connection
with the alternative programs of string theory and loop quantum
gravity. The achievements of the present work include:

1) the analysis of the so-called {\it Hole} phenomenology in
strict connection with the Hamiltonian treatment of the initial
value problem. The work is carried through in metric gravity for
the class of Christoudoulou-Klainermann space-times, in which the
temporal evolution is ruled by the {\it weak} ADM energy. It is
crucial to our analysis the re-interpretation of {\it active}
diffeomorphisms as {\it passive and metric-dependent} dynamical
symmetries of Einstein's equations, a re-interpretation which
enables to disclose their (nearly unknown) connection to gauge
transformations on-shell; this is expounded in the first paper
(gr-qc/0403081).

2) the utilization of the Bergmann-Komar {\it intrinsic
pseudo-coordinates}, defined as suitable functionals of the Weyl
curvature scalars, as tools for a peculiar gauge-fixing to the
super-hamiltonian and super-momentum constraints;

3) the consequent construction of a {\it physical atlas} of
4-coordinate systems for the 4-dimensional {\it mathematical}
manifold, in terms of the highly non-local degrees of freedom of
the gravitational field (its four independent {\it Dirac
observables}). Such construction embodies the {\it physical
individuation} of the points of space-time as {\it point-events},
both in absence and presence of matter, and associates a {\it
non-commutative structure} to each gauge fixing or 4-dimensional
coordinate system.

4) a clarification of the multiple definition given by Peter
Bergmann  of the concept of {\it (Bergmann) observable} in general
relativity. This clarification leads to the proposal of a {\it
main conjecture} asserting the existence of i) special Dirac's
observables which are also Bergmann's observables, ii) gauge
variables that are coordinate independent (namely they behave like
the tetradic scalar fields of the Newman-Penrose formalism). A
by-product of this achievements is the falsification of a recently
advanced argument  asserting the absence of (any kind of) {\it
change} in the observable quantities of general relativity.

5) a proposal showing how the physical individuation of
point-events could in principle be implemented as an experimental
setup and protocol leading to a {\it standard of space-time} more
or less like atomic clocks define standards of time.

In the end, against the well-known Einstein's assertion according
to which {\it general covariance takes away from space and time
the last remnant of physical objectivity}, we conclude that
point-events maintain a {\it peculiar sort of objectivity}. Also,
besides being operationally essential for building measuring
apparatuses for the gravitational field, the role of matter in the
non-vacuum gravitational case is also that of {\it participating
directly} in the individuation process, being involved in the
determination of the Dirac observables. Finally, some hints
following from our approach for the quantum gravity programme are
suggested.

\today

\end{abstract}

\maketitle

\vfill\eject

\section{Introduction.}

In a first paper\cite{1} (hereafter referred to as I), we have
shown how the capabilities of the ADM Hamiltonian approach to
metric gravity enables us to get new insights into a series of
technical problems concerning the physical status of the gauge
variables and the Dirac observables (DO), as well as the dynamical
nature of the simultaneity and gravito-magnetism {\it
conventions}. We have shown in particular that i) before solving
Einstein-Hamilton equations different {\it conventions} within the
same space-time {\it universe} simply correspond to different
gauge choices; ii) {\it each solution of Einstein-Hamilton
equations dynamically selects a preferred convention}. In the
present paper we exploit the technical achievements obtained in I
to get new insights into issues deeply rooted into the
foundational level of the theory that we deem still worth of
clarification. The superiority of the Hamiltonian treatment is
essentially due to the fact that it allows to work {\it off
shell}, i.e., without immediate restriction to the solution of
Einstein's equations. On the other hand, such kind of analysis
could hardly be dealt with in a satisfactory way within the
standard Lagrangian approach because of the non-hyperbolic nature
of Einstein's equations. It is not by chance that the modern
treatment of the initial value problem within the Lagrangian
configurational approach \cite{2,3} must in fact mimic the
Hamiltonian methods.

The adoption of the Hamiltonian viewpoint entails that the range
of our analysis and conclusions be confined to a particular class
of models of general relativity, namely those which are compatible
with a 3+1 splitting of space-time. In particular, we shall work
with globally-hyperbolic, non-compact, topologically trivial,
asymptotically flat at spatial infinity (and with suitable
boundary conditions there) space-times, which are of the type
classified by Christodoulou and Klainermann\cite{4}.

The main specific issues we want to scrutinize are: a) the
long-standing issue of the objectivity of point-events of
space-time. Although this question is nowadays mainly of interest
to philosophers of science and appears to have been nicely
bypassed in the standard physical literature, we intend to show
that it maintains interesting technical aspects which could even
be relevant for the forefront physics, namely string theory and
loop quantum gravity. b) The concept of observable in general
relativity, in particular the relation between the notion of {\it
Dirac observable} \,(DO) and that of {\it Bergmann observable}
(BO), two notions that do not simply overlap. Actually, we shall
show that the relation between these two concepts contains the
seeds of interesting developments concerning not only the concept
of observable itself but also a possible invariant notion of {\it
generalized inertial effects} in general relativity and thereby
new insights into the equivalence principle.

The paper should be read in sequence after I which contains
various technical premises for the present analysis. Previous
partial accounts of the material of this paper can be found in
Refs. \cite{5,6,7}.

\bigskip

General relativity is commonly thought to imply that space-time
points have no {\it intrinsic physical meaning} due to the general
covariance of Einstein's equations. This feature is implicitly
described in standard modern textbooks by the statement that
solutions to the Einstein's equations related by ({\it active})
diffeomorphisms have physically identical properties, so that only
the equivalence class of such solutions represents a
\emph{space-time geometry}. Such kind of equivalence, which also
embodies the modern understanding of Einstein's historical {\it
Hole Argument}, has been named as {\it Leibniz equivalence} in the
philosophical literature by Earman and Norton\cite{8}. In this
paper we will not examine any philosophical aspect of the
issue\footnote{A philosophical critique following from the
technical results of the present couple of papers can be found in
Ref.\cite{9} }, although our analysis is inspired by the belief
that {\it Leibniz equivalence} is not and cannot be the last word
about the {\it intrinsic physical properties} of space-time, well
beyond the needs of the empirical grounding of the theory.
Specifically, our contribution should be inscribed in the list of
the various attempts made in the literature to gain an intrinsic
{\it dynamical characterization of space-time points in terms of
the gravitational field itself}, besides and beyond the trivial
mathematical individuation furnished to them by the coordinates.
We refer in particular to old hints offered by Synge, and to the
attempts successively sketched by Komar, Bergmann and Stachel.
Actually, we claim that we have pursued this line of thought till
its natural end.

\medskip

The original Hole Argument is naturally spelled out within the
configurational Lagrangian framework of Einstein's theory. It is
essential to realize from the beginning that - by its very
formulation - the {\it Hole Argument} is {\it inextricably
entangled} with the initial value problem although, strangely
enough, it has never been explicitly discussed in that context in
a systematic way. Possibly the reason is that most authors have
implicitly adopted the Lagrangian approach, where the Cauchy
problem is intractable because of the non-hyperbolic nature of
Einstein's equations. The proper way to deal with such problem is
indeed the ADM Hamiltonian framework with its realm of DO and
gauge variables. But then the real difficulty is just the
connection between such different frameworks, particularly from
the point of view of {\it symmetries}. The clarification of this
issue in I started from a rediscovery of a nearly forgotten paper
by Bergmann and Komar \cite{10} which enabled us to enlighten this
correspondence of symmetries and, in particular, that existing
between {\it active diffeomorphisms} of the configurational
approach and {\it gauge transformations} of the Hamiltonian
viewpoint.

\medskip

At first sight it could seem that in facing the original
\emph{Hole Argument} Einstein simply equated general covariance
with the unavoidable {\it arbitrariness of the choice of
coordinates}, a fact that, in modern language, can be translated
into {\it invariance under passive diffeomorphisms}. The so-called
{\it point-coincidence argument} (a terminology introduced by
Stachel in 1980), satisfied Einstein doubts at the end of 1915 but
offered mainly a pragmatic solution of the issue and was based on
a very idealized model of physical measurement where all possible
observations reduce to the intersections of the world-lines of
observers, measuring instruments, and measured physical objects.
Furthermore, this solution left unexplored some important aspects
of the role played by the metric tensor in the Hole Argument as
well as of the related underlying full mathematical structure of
the theory.

That the Hole Argument was in fact a subtler issue that Einstein
seemingly thought in 1915 \cite{11} and that it consisted in much
more than mere arbitrariness in the choice of the
coordinates\footnote{In fact, however, Einstein's argument was not
so naive, see Norton \cite{12}.}, has been revealed by a seminal
talk given by John Stachel in 1980 \cite{13}, which gave new life
to the original Hole Argument.

\medskip

The Hole Argument, in its modern version, runs as follows.
Consider a general-relativistic space-time, as specified by the
four-dimensional mathematical manifold $M^4$ and by a metric
tensor field ${}^4g$ which {\it represents at the same time the
chrono-geometrical and causal structure of space-time and the
potential for the gravitational field}. The metric ${}^4g$ is a
solution of the generally-covariant Einstein equations. If any
non-gravitational physical fields are present, they are
represented by tensor fields that are also dynamical fields, and
that appear as sources in the Einstein equations.

Assume now that $M^4$ contains a {\it Hole} $\mathcal{H}$: that
is, an open region where all the non-gravitational fields are
zero. On $M^4$ we can prescribe an {\it active} diffeomorphism
(see I, Section II) $D_A$ that re-maps the points inside
$\mathcal{H}$, but blends smoothly into the identity map outside
$\mathcal{H}$ and on the boundary. Now, {\it just because
Einstein's equations are generally covariant} so that they can be
written down as {\it geometrical relations}, if ${}^4g$ is one of
their solutions, so is the {\it drag-along} field ${}^4g' = D_A^*
\cdot {}^4g$. By construction, for any point $p \in \mathcal{H}$
we have (geometrically) ${}^4g'(D_A \cdot p) = {}^4g(p)$, but of
course ${}^4g'(p) \neq {}^4g(p)$ (also geometrically). Now, what
is the correct interpretation of the new field ${}^4g'$? Clearly,
the transformation entails an {\it active redistribution of the
metric over the points of the manifold}, so the crucial question
is whether, to what extent, and how the points of the manifold are
primarily {\it individuated}.
\medskip

In the mathematical literature about topological spaces, it is
always implicitly assumed that the entities of the set can be
distinguished and considered separately (provided the Hausdorff
conditions are satisfied), otherwise one could not even talk about
point mappings or homeomorphisms. It is well known, however, that
the points of a homogeneous space cannot have any intrinsic
individuality\footnote{ As Hermann Weyl \cite{14} puts it: ''There
is no distinguishing objective property by which one could tell
apart one point from all others in a homogeneous space: at this
level, fixation of a point is possible only by a {\it
demonstrative act} as indicated by terms like {\it this} and {\it
there}.''}. There is only one way to individuate points at the
mathematical level that we are considering: namely by {\it
coordinatization}, a procedure that transfers the individuality of
$4$-tuples of real numbers to the elements of the topological set.
Precisely, we introduce by convention {\it a standard} coordinate
system for the {\it primary individuation} of the points (like the
choice of {\it standards} in metrology). Then, we can get as many
different {\it names}, for what we consider the same primary
individuation, as the coordinate charts containing the point in
the chosen atlas of the manifold. We can say, therefore, that all
the relevant transformations operated on the manifold $M^4$
(including {\it active} diffeomorphisms which map points to
points), even if viewed in purely geometrical terms, {\it must} be
realizable in terms of (possibly generalized) coordinate
transformations.

Let us go back to the effect of this {\it primary} mathematical
individuation of manifold points. If we now think of the points of
$\mathcal{H}$ as also {\it physically individuated}
spatio-temporal events even before the metric is defined, then
${}^4g$ and ${}^4g'$ must be regarded as {\it physically distinct}
solutions of the Einstein equations (after all, as already noted,
${}^4g'(p) \neq {}^4g(p)$ at the {\it same} point $p$). This,
however, is a devastating conclusion for the causality, or better,
{\it determinateness} of the theory, because it implies that, even
after we completely specify a physical solution for the
gravitational and non-gravitational fields outside the Hole - for
example, on a Cauchy surface for the initial value problem,
assuming for the sake of argument that this intuitive and
qualitative wording is mathematically correct, see Section III -
we are still {\it unable to predict uniquely the physical solution
within the Hole}. Clearly, if general relativity has to make any
sense as a physical theory, there must be a way out of this
foundational quandary, {\it independently of any philosophical
consideration}.

In the modern understanding, the most widely embraced escape from
the (mathematical) strictures of the Hole Argument (which is
essentially an update to current mathematical terms of the
pragmatic solution adopted by Einstein), is to {\it deny that
diffeomorphically related  mathematical solutions represent
physically distinct solutions}. With this assumption (i.e., the
mathematical basis of Leibnitz equivalence), {\it an entire
equivalence class of diffeomorphically related mathematical
solutions represents only one physical solution}. This statement,
is implicitly taken as obvious in the contemporary specialized
literature (see, e.g. Ref.\cite{15}).

It is seen at this point that the conceptual content of general
covariance is far more deeper than the simple invariance under
arbitrary changes of coordinates. Actually (see Stachel
\cite{16,17}) asserting that ${}^4g$ and $D_A^* \cdot {}^4g$
represent {\it one and the same gravitational field} entails that
{\it the mathematical individuation of the points of the
differentiable manifold by their coordinates has no physical
content until a metric tensor is specified}. In particular,
coordinates lose any {\it physical significance whatsoever}
\cite{11}. Furthermore, if ${}^4g$ and $D_A^* \cdot {}^4g$ must
represent the same gravitational field, they cannot be physically
distinguishable in any way. So when we act on ${}^4g$ with an
active diffeomorphisms to create the drag-along field $D_A^* \cdot
{}^4g$, {\it no element of physical significance} can be left
behind: in particular, nothing that could identify a point $p$ of
the manifold as the {\it same} point of space-time for both
${}^4g$ and $D_A^* \cdot {}^4g$. Instead, when $p$ is mapped onto
$p' = D_A \cdot p$, it {\it brings over its identity}, as
specified by ${}^4g'(p')= {}^4g(p)$ \footnote{A further important
point made by Stachel is that simply because a theory has
generally covariant equations, it does not follow that the points
of the underlying manifold must lack any kind of physical
individuation. Indeed, what really matters is that there can be no
\emph{non-dynamical individuating field} that is specified
\emph{independently} of the dynamical fields, and in particular
independently of the metric. If this was the case, a
\emph{relative} drag-along of the metric with respect to the
(supposedly) individuating field would be physically significant
and would generate an inescapable Hole problem. Thus, the absence
of any non-dynamical individuating field, as well as of any
dynamical individuating field independent of the metric, is the
crucial feature of the purely gravitational solutions of general
relativity as well as of the very {\it concept} of {\it general
covariance}.}.

This conclusion led Stachel to the conviction that space-time
points {\it must} be {\it physically} individuated {\it before}
space-time itself acquires a physical bearing, and that the metric
itself plays the privileged role of {\it individuating field}: a
necessarily {\it unique role} in the case of space-time {\it
without matter}. More precisely, Stachel claimed that this
individuating role should be implemented by four invariant
functionals of the metric, already considered by Bergmann and
Komar \cite{18} (see Section II). However, he did not follow up on
his suggestion. As a matter of fact, as we shall see, the question
is not straightforward.
\medskip

There are many reasons why one should revisit the Hole Argument
nowadays, quite apart from any {\it conceptual} interest.
\medskip

First of all, the crucial point of the Hole issue is that the
mathematical representation of space-time provided by general
relativity under the condition of general covariance evidently
contains {\it superfluous structure} hidden behind Leibniz
equivalence and that this structure must be isolated. At the level
of general covariance, only the equivalence class is physically
real so that, on this understanding, general covariance is
invariably an unbroken symmetry and the physical world is to be
described in a diffeomorphically invariant way. Of course, the
price to be paid is that the values of all fields at manifold
points as specified by the coordinates, are not physically real.
On the other hand, this isolation appears to be required {\it de
facto} both by any explicit solution of Einstein's equation, which
requires specification of the arbitrariness of coordinates, and by
the empirical foundation of the theory: after all any effective
kind of measurement requires in fact a definite physical
individuation of space-time points in terms of physically
meaningful coordinates. Summarizing, it is evident that breaking
general covariance is a pre-condition for the isolation of the
superfluous structure hidden within {\it Leibniz equivalence},
namely the generalized inertial effects analyzed in I.

\medskip

Secondly, the program of the physical individuation of space-time
points must be completed because, as it will appear evident in
Section II, the mere recourse to the four functional invariants of
the metric alluded to by Stachel cannot do, by itself, the job of
physically individuating space-time points. In the context of the
Hamiltonian formalism, we find the tools for completing Stachel's
suggestion and exploiting the old proposal advanced by Bergmann
and Komar for an intrinsic labeling of space-time points by means
of the eigenvalues of the Weyl tensor. Precisely, Bergman and
Komar, in a series of papers \cite{18,19,20} introduced suitable
invariant scalar functionals of the metric and its first
derivatives as {\it invariant pseudo-coordinates}\footnote{
Actually, the first suggestion of specifying space-time points
{\it absolutely} in terms of curvature invariants is due to Synge
\cite{21} b)}. We shall show that such proposal can be utilized in
constructing a peculiar {\it gauge-fixing to the super-hamiltonian
and super-momentum constraints} in the canonical reduction of
general relativity. This gauge-fixing makes the {\it invariant
pseudo-coordinates} into effective {\it individuating fields} by
forcing them to be {\it numerically} identical with ordinary
coordinates: in this way the individuating fields turn the {\it
mathematical} points of space-time into {\it physical
point-events}. Eventually, we discover that what really
individuates space-time points physically are the very {\it
degrees of freedom of the gravitational field}. As a
consequence,\- we advance the claim that - physically - Einstein's
vacuum space-time is literally {\it identified} with the
autonomous physical degrees of freedom of the gravitational field,
while the specific functional form of the {\it invariant
pseudo-coordinates} matches these latter into the manifold's
points. The introduction of matter has the effect of modifying the
Riemann and Weyl tensors, namely the curvature of the
4-dimensional substratum, and to allow {\it measuring} the
gravitational field in a geometric way for instance through
effects like the geodesic deviation equation. It is important to
emphasize, however, that the addition of {\it matter} does not
modify the construction leading to the individuation of
point-events, rather it makes it {\it conceptually more
appealing}.
\medskip

Finally, our procedure of individuation transfers, as it were, the
noncommutative Poisson-Dirac structure of the DO onto the
individuated point-events. The physical implications of this
circumstance might deserve some attention in view of the
quantization of general relativity. Some hints for the quantum
gravity programme will be offered in the final Section of the
paper (Concluding Remarks).

\bigskip

A Section of the paper is devoted to our second main topic: the
clarification of the {\it multiple} and rather ambiguous concept
of {\it Bergmann's observable} (BO) \cite{22}. Bergmann's
definition has various facets, namely a {\it configurational} side
having to do with invariance under {\it passive} diffeomorphisms,
an {\it Hamiltonian} side having to do with Dirac's concept of
observable, and the property of {\it predictability} which is
entangled with both sides. According to Bergmann, (his) {\it
observables} are passive diffeomorphisms invariant quantities
(PDIQ) "which can be predicted uniquely from initial data", or
"quantities that are invariant under a coordinate transformation
that leaves the initial data unchanged". Bergmann says in addition
that they are further required to be gauge invariant, a statement
that could only be interpreted as implying that Bergmann's
observables are simultaneously DO. Yet, he offers no explicit
demonstration of the compatibility of this bundle of statements.

Actually, once fully clarified, the concept of {\it
predictability} implies in its turn that, in order Bergmann's {\it
multiple definition} be consistent, {\it only four} of such
observables can exist for the vacuum gravitational field, and can
be nothing else than tensorial Lagrangian counterparts of the
Hamiltonian DO. We formalize this result and related consequences
into a {\it main conjecture}, which essentially amounts to
claiming the internal consistency of Bergmann's {\it multiple
definition}. Our conjecture asserts: i) the existence of special
Dirac's observables which are also Bergmann's observables, as well
as ii) the existence of gauge variables that are coordinate
independent.

\bigskip \noindent

As anticipated in I, the hoped for validity of the {\it main
conjecture} would add new emphasis to the physical meaning of the
separation between DO, as related to {\it tidal-like} effects, on
the one hand, and gauge variables, as related to {\it generalized
inertial} effects, on the other. Actually, in spite of the
physical relevance of distinction as it stands, its weakness is
that the separation of the two autonomous degrees of freedom of
the gravitational field from the gauge variables is, as yet, a
coordinate (i.e. gauge) - dependent concept. The known examples of
pairs of conjugate DO are neither coordinate-independent (they are
not PDIQ) nor tensors. Bergmann asserts that the only known method
(at the time) to build BO is based on the existence of
Bergmann-Komar invariant pseudo-coordinates. The results of this
method, however, are of difficult interpretation, so that, in
spite of the importance of this alternative non-Hamiltonian
definition of observables, no explicit determination of them has
been proposed so far. A possible starting point to attack the
problem of the connection of DO with BO seems to be a Hamiltonian
re-formulation of the Newman-Penrose formalism \cite{23} (it
contains only PDIQ) employing Hamiltonian null-tetrads carried by
the time-like observers of the congruence orthogonal to the
admissible space-like hyper-surfaces. This suggests the technical
{\it conjecture} that special Darboux bases for canonical gravity
should exist in which generalized inertial effects (related to
gauge variables fixings) are described by PDIQ while the
autonomous degrees of freedom (DO) are {\it also} BO. Therefore,
the validity of the conjecture would render our distinction above
an invariant statement, providing a remarkable contribution to the
old-standing debate about the equivalence principle. Note in
addition that, since Newman-Penrose PDIQ are tetradic quantities,
the validity of the conjecture would also eliminate the existing
difference between the observables for the gravitational field and
the observables for matter, built usually by means of the tetrads
associated to some time-like observer. Furthermore, this would
also provide a starting point for defining a metrology in general
relativity in a generally covariant way\footnote{Recall that this
is the main conceptual difference from the non-dynamical metrology
of special relativity}, replacing the empirical metrology
\cite{24} used till now. It would also enable to replace the {\it
test matter} of the axiomatic approach to measurement theory by
dynamical matter (see Appendix A).

\medskip

Incidentally, our results about the definition of {\it Bergmann
observable} help in showing that various recent claims \cite{25}
about the absence of any kind of {\it change} in general
relativity \emph{as such} are not mathematically justified and
that - with reference to the class of models we are considering -
the so-called issue of time raises no particular difficulty: even
more, such models provide an explicit {\it counterexample} to the
{\it frozen time argument}. The role of the generator of real time
evolution in such space-times is played in fact by the so-called
{\it weak} ADM energy, while the super-hamiltonian constraint has
nothing to do with temporal change and is only the generator of
gauge transformations connecting different admissible 3+1
splittings of space-time. We argue, therefore, that in these
space-times there is neither a frozen reduced phase space nor a
possible Wheeler-De Witt interpretation based on some local
concept of time as in compact space-times. In conclusion, we claim
that our gauge-invariant approach to general relativity is {\it
perfectly adequate to accommodate real temporal change}, so that
all the consequent developments based on it are immune to
criticisms like those referred to above.

\medskip

A final step of our analysis consists in suggesting how the
physical individuation of space-time points, introduced at the
conceptual level, could in principle be implemented with {\it a
well-defined empirical procedure, an experimental set-up and
protocol for positioning and orientation}. This suggestion is
outlined in correspondence with the abstract treatment of the
empirical foundation of general relativity as exposed in the
classical paper of Ehlers, Pirani and Schild \cite{26}. The
conjunction of the Hamiltonian treatment of the initial value
problem, with the correlated physical individuation of space-time
points, and the practice of general-relativistic measurement, on
the backdrop of the axiomatic foundation closes, as it were, the
{\it coordinative circuit} of general relativity.
\bigskip

Section IIA is devoted to the issue of the individuation of the
{\it mathematical points} of $M^{4}$ as {\it physical
point-events} by means of a peculiar gauge-fixing to
Bergmann-Komar {\it intrinsic pseudo-coordinates}. In Section IIB,
we sketch the implementation of the physical individuation in
terms of well-defined experimental procedures which realize the
axiomatic structure of general relativity proposed by Ehlers,
Pirani and Schild. The analysis of the concept of BO and the
criticism of the {\it frozen time} and {\it universal no-change}
arguments are the content of Section III where our {\it main
conjecture} is advanced concerning the relations between DO and
BO. The Conclusion contains some general comments about the gauge
nature of general relativity and some hints in view of the quantum
gravity programme while Appendix A reviews the Ehlers, Pirani and
Schild axiomatic approach.

\vfill\eject

\section{Physical Individuation of Space-Time Points by means of
Gauge Fixings to Bergmann-Komar Intrinsic Coordinates.}

Let us now exploit the results of I, Sections II and III, to the
effect of clarifying the issue of the physical individuality of
space-time point-events in general relativity and its implications
for the theory of measurements with test objects.

\subsection{The Physical Individuation of Space-Time Points.}

Let us begin by recalling again that the ADM formulation assumes
the existence of a mathematical 4-manifold, the space-time $M^4$,
admitting 3+1 splittings with space-like leaves
$\Sigma_{\tau}\approx R^3$. All fields (also matter fields when
present) depend on $\Sigma_{\tau}$-adapted coordinates $(\tau
,\vec \sigma )$ for $M^4$.

We must insist again that a crucial component of the individuation
issue is the inextricable entanglement of the Hole Argument with
the initial value problem, which has been dealt with at length in
I. Now, however, we have at our disposal the right framework for
dealing with the initial value problem, so our main task should be
to put all things together. Finally, another fundamental tool at
our disposal is the clarification we gained in Section II of I
concerning the relation between active diffeomorphisms in their
passive view and the dynamical gauge symmetries of Einstein's
equations in the Hamiltonian approach.

We are then ready to move forward by conjoining Stachel's
suggestion with the proposal advanced by Bergmann and Komar
\cite{18} that, in the absence of matter fields, the values of
four invariant scalar fields built from the contractions of the
Weyl tensor (actually its eigenvalues) can be used to build  {\it
intrinsic pseudo-coordinates}\footnote{As shown in Ref.\cite{27}
in general space-times with matter there are 14 algebraically
independent curvature scalars for $M^4$.}.

The four invariant scalar eigenvalues $\Lambda^{(k)}_W(\tau ,\vec
\sigma )$, $k=1,..,4$, of the Weyl tensor, written in Petrov
compressed notations, are

\bea
 &&\Lambda^{(1)}_W = Tr\, ({}^4C\, {}^4g\, {}^4C\,
 {}^4g),\nonumber \\
 &&\Lambda^{(2)}_W = Tr\, ({}^4C\, {}^4g\, {}^4C\, {}^4\epsilon ),
 \nonumber \\
 &&\Lambda^{(3)}_W = Tr\, ({}^4C\, {}^4g\, {}^4C\, {}^4g\, {}^4C\,
 {}^4g),\nonumber \\
 &&\Lambda^{(4)}_W = Tr\, ({}^4C\, {}^4g\, {}^4C\, {}^4g\, {}^4C\,
 {}^4\epsilon ),
 \label{IV1}
 \eea

\noindent where ${}^4C$ is the Weyl tensor, $^4g$ the metric, and
${}^4\epsilon$ the Levi-Civita totally anti-symmetric tensor.

Bergmann and Komar then propose that we build a set of (off-shell)
{\it intrinsic coordinates} for the point-events of space-time as
four suitable functions of the $\Lambda^{(k)}_W$'s,

\beq
 {\bar \sigma}^{\bar A}(\sigma ) = F^{\bar A}[\Lambda^{(k)}_W
 [{}^4g(\sigma), \partial\, {}^4g(\sigma)]],
\,(\bar A = 1,2,...,4).
 \label{IV2}
  \eeq

\noindent Indeed, under the hypothesis of no space-time
symmetries,\footnote{Our attempt to use intrinsic coordinates to
provide a physical individuation of point-events would {\it prima
facie} fail in the presence of symmetries (with or without
matter), when the $F^{\bar A}[\Lambda^{(k)}_W [{}^4g(\sigma ),
\partial\, {}^4g(\sigma )]]$ become degenerate. This objection was
originally raised by Norton \cite{28} as a critique to
manifold-plus-further-structure (MPFS) substantivalism (according
to which the points of the manifold, conjoined with additional
local structure such as the metric field, can be considered
physically real; see for instance \cite{29}). Several responses
are possible. First, although most of  the {\it known} exact
solutions of the Einstein equations admit one or more symmetries,
these mathematical models are very idealized and simplified; in a
realistic situation (for instance, even with two masses)
space-time is filled with the excitations of the gravitational
degrees of freedom, and admits no symmetries at all. A case study
is furnished by the non-symmetric and non-singular space-times of
Christodoulou-Klainermann \cite{4}. Second, the parameters of the
symmetry transformations can be used as supplementary
individuating fields, since, as noticed by Komar \cite{20} and
Stachel \cite{17} they also depend upon metric field, through its
isometries. To this move it has been objected \cite{30} that these
parameters are purely mathematical artifacts, but a simple
rejoinder is that the symmetric models too are mathematical
artifacts. Third, and most important, in our analysis of the
physical individuation of points we are arguing a question of
principle, and therefore we must consider {\it generic} solutions
of the Einstein equations rather than the null-measure set of
solutions with symmetries.} we would be tempted (like Stachel) to
use the $F^{\bar A}[\Lambda^{(k)}_W]$ as individuating fields to
{\it label the points of space-time}, at least
locally.\footnote{Problems might arise if we try to extend the
labels to the entire space-time: for instance, the coordinates
might turn out to be multi-valued.}

Of course, since they are invariant functionals, the $F^{\bar
A}[\Lambda^{(k)}_W]$'s are quantities invariant under passive
diffeomorphisms (PDIQ), therefore, as such, they do not define a
coordinate chart for the atlas of the mathematical Riemannian
4-manifold $M^4$ in the usual sense (hence the name of {\it
pseudo-coordinates} and the superior bar we used in $F^{\bar A}$).
Moreover, the tetradic 4-metric which can be built by means of the
intrinsic pseudo-coordinates (see the next Section) is a formal
object invariant under passive diffeomorphisms that does not
satisfy Einstein's equations (but possibly much more complex
derived equations). Therefore, the action of active
diffeomorphisms on the tetradic metric is not {\it directly}
connected to the Hole argument. All this leads to the conclusion
that the proposal advanced by Bergmann \cite{22} ("we might then
identify a {\it world point} (location-plus-instant-in-time) by
the values assumed by (the four intrinsic pseudo-coordinates)") to
the effect of individuating point-events in terms of {\it
intrinsic pseudo-coordinates} is not - as it stands - {\it
physically viable} in a tractable way. This is not the final
verdict, however, and we must find a dynamical bridge between the
intrinsic pseudo-coordinates and the ordinary 4-coordinate systems
which provide the primary identification of the points of the
mathematical manifold.
\medskip

Our procedure starts when we recall that, within the Hamiltonian
approach, Bergmann and Komar \cite{18} proved the fundamental
result that the Weyl eigenvalues $\Lambda^{(A)}_W$, once
re-expressed as functionals of the  ADM canonical variables, {\it
do not depend on the lapse and shift functions} but only on the
3-metric and its conjugate canonical momentum,
$\Lambda^{(k)}_W[{}^4g(\tau ,\vec \sigma),
\partial\, {}^4g(\tau ,\vec \sigma)] = {\tilde \Lambda}^{(k)}_W[{}^3g(\tau
,\vec \sigma), {}^3\Pi(\tau ,\vec \sigma)]$. This result entails
that the {\it intrinsic pseudo-coordinates} ${\bar \sigma}^{\bar
A}$ are natural quantities to be used to build four gauge fixing
constraints for the canonical reduction procedure depending only
on a single hyper-surface $\Sigma_{\tau}$ and not on how these
surfaces are packed in the foliation.
\medskip

Taking into account the results of Section III of I, we know that,
in a completely fixed  gauge, both the four intrinsic {\it
pseudo-coordinates} and the ten {\it tetradic} components of the
metric field (see Eq.(\ref{V2}) of the next Section)  become gauge
dependent functions of the four DO of that gauge. For the Weyl
scalars in particular we can write:

\beq
 \Lambda^{(k)}_W(\tau ,\vec \sigma ){|}_G =  {\tilde \Lambda}^{(k)}_W[{}^3g(\tau
,\vec \sigma), {}^3\Pi(\tau ,\vec \sigma)]{|}_G =
\Lambda_G^{(k)}[r^G_{\bar a}(\tau ,\vec \sigma ), \pi^G_{\bar
a}(\tau ,\vec \sigma )].
 \label{IV3}
  \eeq

\noindent where ${|}_G$ denotes the specific gauge. Conversely, by
the inverse function theorem, in each gauge, the DO of that gauge
can be expressed as functions of the 4 eigenvalues restricted to
that gauge: $\Lambda^{(k)}_W (\tau ,\vec \sigma ){|}_G $.

\bigskip

Our program is implemented in the following way: after having
selected a {\it completely arbitrary mathematical} radar-type (see
Ref.\cite{31}) coordinate system $\sigma^A \equiv [\tau,\sigma^a]$
adapted to the $\Sigma_\tau$ surfaces, we choose {\it as physical
individuating fields} four suitable functions $F^{\bar
A}[\Lambda^{(k)}_W(\tau ,\vec \sigma )]$, and express them as
functionals $\tilde F^{\bar A}$ of the ADM variables

\beq
 F^{\bar A}[\Lambda^{(k)}_W(\tau ,\vec \sigma )] = F^{\bar
A}[{\tilde \Lambda}^{(k)}_W[{}^3g(\tau ,\vec \sigma), {}^3\Pi(\tau
,\vec \sigma)]] = \tilde F^{\bar A}[{}^3g(\tau ,\vec \sigma),
{}^3\Pi(\tau ,\vec \sigma)].
 \label{IV4}
  \eeq

\noindent The space-time points, {\it mathematically individuated}
by the quadruples of real numbers $\sigma^A$, become now {\it
physically individuated point-events} through the imposition of
the following gauge fixings to the four secondary constraints

\beq
 {\bar \chi}^A(\tau ,\vec \sigma )\- {\buildrel {def} \over =}\-  \sigma^A -
 {\bar \sigma}^{\bar A}(\tau ,\vec \sigma ) = \sigma^A -
 F^{\bar A}\Big[{\tilde \Lambda}^{(k)}_W[{}^3g(\tau ,\vec \sigma), {}^3\Pi(\tau ,\vec
\sigma)]\Big] \approx 0.
 \label{IV5}
 \eeq

\noindent Then, following the standard procedure, we end with a
completely fixed Hamiltonian gauge, say $G$. This will be a
correct gauge fixing provided the functions $F^{\bar A}[{
\Lambda}^{(k)}_W(\tau ,\vec \sigma )]$ are chosen so that the
${\bar \chi}^A(\tau ,\vec \sigma )$'s satisfy the {\it orbit
conditions}

\beq
 det\, | \{ {\bar \chi}^A(\tau ,\vec \sigma ), {\tilde {\cal
 H}}^B(\tau ,{\vec \sigma}^{'}) \} | \not= 0,
 \label{IV6}
 \eeq

\noindent where ${\tilde {\cal H}}^B(\tau ,\vec \sigma ) = \Big(
{\tilde {\cal H}}(\tau ,\vec \sigma ); {}^3{\tilde {\cal
H}}^r(\tau ,\vec \sigma ) \Big) \approx 0$ are the
super-hamiltonian and super-momentum constraints of Eqs.(3.2) of
I. These conditions enforce the Lorentz signature on
Eq.(\ref{IV5}), namely the requirement that $F^{\bar \tau}$ be a
{\it time} variable, and imply that {\it the $F^{\bar A}$'s cannot
be DO}.

The above gauge fixings allow in turn the determination of the
four Hamiltonian gauge variables $\xi^r(\tau ,\vec \sigma )$,
$\pi_{\phi}(\tau ,\vec \sigma )$ of Eqs.(3.7) of I. Then, their
time constancy induces the further gauge fixings ${\bar
\psi}^A(\tau ,\vec \sigma ) \approx 0$ for the determination of
the remaining gauge variables, i.e., the lapse and shift functions
in terms of the DO in that gauge as

\bea
 {\dot {\bar \chi}}^A(\tau ,\vec \sigma ) &=& {{\partial {\bar
 \chi}^A(\tau ,\vec \sigma )} \over {\partial \tau}} + \{
 {\bar \sigma}^{\bar A}(\tau ,\vec \sigma ), {\bar H}_D \} = \delta^{A\tau}
 +\nonumber \\
 &+& \int d^3\sigma_1\, \Big[ n(\tau ,{\vec \sigma}_1)\, \{
  \sigma^{\bar A}(\tau ,\vec \sigma ), {\cal H}(\tau ,{\vec
  \sigma}_1) \} + n_r(\tau ,{\vec \sigma}_1)\, \{  \sigma^{\bar A}(\tau ,\vec \sigma ),
  {\cal H}^r(\tau ,{\vec \sigma}_1) \}\Big] =\nonumber \\
  &=& {\bar \psi}^A(\tau ,\vec \sigma ) \approx 0.
  \label{IV7}
  \eea

\noindent Finally, ${\dot {\bar \psi}}^A(\tau ,\vec \sigma )
\approx 0$ determines the Dirac multipliers $\lambda^A(\tau ,\vec
\sigma )$.\bigskip

In conclusion, the gauge fixings (\ref{IV5}) ({\it which break
general covariance}) constitute the crucial bridge that transforms
the {\it intrinsic pseudo-coordinates} into {\it true physical
individuating coordinates}. \medskip

As a matter of fact, after going to Dirac brackets, we enforce the
point-events individuation in the form of the {\it identity}

\beq
 \sigma^A \equiv {\bar \sigma}^{\bar A} = {\tilde F}^{\bar A}_{G}[
r^G_{\bar a}(\tau ,\vec \sigma ), \pi^G_{\bar a}(\tau , \vec
\sigma)] = F^{\bar A}[\Lambda^{(k)}_W(\tau ,\vec \sigma )]{|}_G.
 \label{IV8}
  \eeq

\medskip

In this {\it physical 4-coordinate grid}, the 4-metric, as well as
other  fundamental physical entities, like e.g. the space-time
interval $ds^{2}$ with its associated causal structure, and the
lapse and shift functions, depend entirely on the DO in that
gauge. The same is true, in particular, for the solutions of the
eikonal equation \cite{4} ${}^4g^{AB}(\sigma^D)\, {{\partial
U(\sigma^D)}\over {\partial \sigma^A}}\,  {{\partial
U(\sigma^D)}\over {\partial \sigma^B}} = 0$, which define
generalized wave fronts and, therefore, through the envelope of
the null surfaces $U(\sigma^D) = const.$ at a point,  the light
cone at that point.
\bigskip

Let us stress that, according to the results of I, only on the
solutions of Einstein's equations the completely fixed gauge $G$
is equivalent to the fixation of a definite 4-coordinate system
$\sigma^A_G$. Our gauge fixing (\ref{IV5}) ensures that {\it
on-shell} we get $\sigma^A = \sigma^A_G$. In this way we get a
physical 4-coordinate grid on the mathematical 4-manifold $M^4$
dynamically determined by tensors over $M^4$ with a rule which is
invariant under ${}_PDiff\, M^4$ but such that the functional form
of the map {\it $\sigma^A \mapsto \, physical\,\,\,
4-coordinates$} depends on the complete chosen gauge $G$: we see
that what is usually called the {\it local point of view}
\cite{32} (see later on) is justified a posteriori in every
completely fixed gauge.

\bigskip

Summarizing, the effect of the whole procedure is that {\it the
values of the {\it DO}, whose dependence on space (and on {\it
parameter} time) is indexed by the chosen radar coordinates $(\tau
,\vec \sigma )$, reproduces precisely such $(\tau ,\vec \sigma )$
as the Bergmann-Komar {\it intrinsic coordinates} in the chosen
gauge $G$}. In this way {\it mathematical points have become
physical individuated point-events} by means of the highly
non-local structure of the DO. If we read the identity (\ref{IV8})
as $\sigma^A \equiv f^{\bar A}_G(r^G_{\bar a}, \pi^G_{\bar a})$,
we see that each coordinate system $\sigma^A$ is determined
on-shell by the values of the 4 canonical degrees of freedom of
the gravitational field in that gauge. This is tantamount to
claiming that {\it the physical role and content of the
gravitational field in absence of matter is just the very
identification of the points of Einstein space-times into physical
point-events by means of its four independent phase space degrees
of freedom}. The existence of physical point-events in general
relativity appears here as a synonym of the existence of the DO,
i.e. of the true physical degrees of freedom of the gravitational
field.
\medskip

As said in the Introduction, the addition of matter does not
change this conclusion, because we can continue to use the gauge
fixing (\ref{IV5}). However, matter changes the Weyl tensor
through Einstein's equations and contributes to the separation of
gauge variables from DO in the quasi-Shanmugadhasan canonical
transformation through the presence of its own DO. In this case we
have DO both for the gravitational field and for the matter
fields, which satisfy coupled Hamilton equations. Therefore, since
the gravitational DO will still provide the individuating fields
for point-events according to our procedure, {\it matter will come
to influence - on-shell only - the very physical individuation of
points}.

\hfill\break

We have seen that, once the orbit conditions are satisfied, the
Bergmann-Komar {\it intrinsic pseudo-coordinates} $F^{\bar
A}[{\tilde \Lambda}^{(k)}_W[{}^3g(\tau ,\vec \sigma), {}^3\Pi(\tau
,\vec \sigma)]{|}_G$ become just the {\it individuating fields}
Stachel was looking for. Indeed, by construction, the intrinsic
pseudo-coordinates are both invariant under ${}_PDiff\, M^4$ and
also {\it numerically invariant} under the drag along induced by
active diffeomorphisms (in the notations of the Introduction we
have $[\phi^*F^{\bar A}](p) \equiv [F^{\bar A\, '}](p) = [F^{\bar
A}](\phi^{-1} \cdot p)\,\,$), a fact that is also essential for
maintaining a connection to the {\it Hole Argument}.
\medskip

A better understanding of our point of view can be achieved by
exploiting Bergmann-Komar's group of passive transformations $Q$
discussed in Section II of I. We can argue in the following way.
Given a 4-coordinate system $\sigma^A$, the passive view of each
active diffeomorphism $\phi$ defines a new 4-coordinate system
$\sigma^A_{\phi}$ (drag-along coordinates produced by a
generalized Bergmann-Komar transformation (2.4) of I). This means
that there will be two functions $F^{\bar A}$ and $F^{\bar
A}_{\phi}$ realizing these two coordinates systems through the
gauge fixings

\bea
 &&\sigma^A - F^{\bar A}[\Lambda^{(k)}_W(\sigma)] \approx
 0,\nonumber \\
 &&\sigma_{\phi}^A - F_{\phi}^{\bar A}[\Lambda^{(k)}_W(\sigma_{\phi})] \approx
 0,
 \label{IV9}
 \eea

It is explicitly seen in this way that the functional freedom in
the choice of the four functions $F^{\bar A}$ allows to cover all
those coordinates charts $\sigma^A$ in the atlas of the
mathematical space-time $M^4$ which are adapted to any allowed 3+1
splitting. By using gauge fixing constraints more general than
those in Eq.(\ref{IV5}) (like the standard gauge fixings used in
ADM metric gravity) we can reach all the 4-coordinates systems of
$M^4$. Here, however, we wanted to restrict to the class of gauge
fixings (\ref{IV5}) for the sake of clarifying the
interpretational issues. \hfill\break

Let us conclude by noting that the gauge fixings (\ref{IV5}),
(\ref{IV7}) induce a {\it coordinate-dependent non-commutative
Poisson bracket structure} upon the {\it physical point-events} of
space-time by means of the associated Dirac brackets implying
Eqs.(\ref{IV8}). More exactly, on-shell, each coordinate system
gets a well defined non-commutative structure determined by the
associated functions ${\tilde F}^{\bar A}_G(r^G_{\bar a},
\pi^G_{\bar a})$, for which we have $\{ {\tilde F}^{\bar
A}_G(r^G_{\bar a}(\tau , \vec \sigma ), \pi^G_{\bar a}(\tau ,\vec
\sigma )), {\tilde F}^{\bar B}_G(r^G_{\bar a}(\tau ,{\vec
\sigma}_1), \pi^G_{\bar a}(\tau ,{\vec \sigma}_1)) \}^* \not= 0$.
The meaning of this structure in view of quantization is worth
investigating (see the Concluding Remarks).\hfill\break

\subsection{Implementing the Physical Individuation of Point-Events
with Well-Defined Empirical Procedures: a Realization of the
Axiomatic Structure of Ehlers, Pirani and Schild.}

The problem of the individuation of space-time points as
point-events cannot be methodologically separated from the problem
of defining a theory of measurement consistent with general
covariance. This means that we should not employ the absolute
chrono-geometric structures of special relativity, like it happens
in all the formulations on a given background (gravitational waves
as a spin two field over Minkowski space-time, string theory,...).
Moreover matter (either test or dynamical) is now an essential
ingredient for defining the experimental setup.

At present we do not have such a theory, but only preliminary
attempts and an empirical metrology \cite{24}, in which the
standard unit of time is a {\it coordinate time} and not a proper
time. As already said, a {\it global non-inertial space-time
laboratory} with its standards corresponds to a description
realized by a completely fixed Hamiltonian gauge viz., being
on-shell, in an atlas of uniquely determined 4-coordinate systems.

\medskip

\noindent We shall take into account the following pieces of
knowledge.\medskip

A) Ehlers, Pirani and Schild \cite{26} developed an axiomatic
framework for the foundations of general relativity and
measurements (reviewed in Appendix A). These authors exploit the
notions of {\it test objects} as {\it idealizations} to the effect
of approximating the {\it conformal, projective, affine} and {\it
metric} structures of Lorentzian manifolds; such structures are
then used to define {ideal geodesic clocks} \cite{33}. The
axiomatic structure refers to basic objects such as {\it test
light rays} and {\it freely falling test particles}. The first
ones are used in principle to reveal the {\it conformal} structure
of space-time, the second ones the {\it projective} structure.
Under an axiom of compatibility which is well corroborated by
experiment (see Ref.\cite{34}), it can be shown that these two
independent classes of observations determine completely the
structure of space-time. Let us remark that one should extend this
axiomatic theory to tetrad gravity (space-times with frames) in
order to include objects like {\it test gyroscopes} needed to
detect gravito-magnetic effects.\footnote{Stachel \cite{35},
stresses the dynamical (not axiomatic) aspect of the general
relativistic space-times structures associated to the behavior of
ideal measuring rods ({\it geometry}) and clocks ({\it
chronometry}) and free test particles ({\it inertial
structures})}.

B) De Witt \cite{36} introduced a procedure for measuring the
gravitational field based on a reference fluid (a stiff elastic
medium) equipped with material clocks. This phenomenological test-
fluid is then exploited to bring in Bergmann-Komar invariant
pseudo-coordinates $\zeta^{\bar A}$, $\bar A =1,..,4$, as a method
for coordinatizing the space-time where to do measurements and
also for grounding space-time geometry operationally, at least in
the weak field regime. De Witt essentially proposes to simulate a
mesh of local clocks and rods. Even if De Witt considers the
measurement of a weak quantum gravitational field smeared over
such a region, his procedure could even be adopted classically. In
this perspective, our approach furnishes the ingredients of the
Hamiltonian description of the gravitational field, which were
lacking at the time De Witt developed his preferred covariant
approach.

C) Antennas and interferometers are the tools used to detect
gravitational waves on the Earth. The mechanical prototype of
these measurements are test springs with end masses feeling the
gravitational field as the tidal effect described by the geodesic
deviation equation \cite{33,37}. Usually, however, one works on
the Minkowski background in the limit of weak field and
non-relativistic velocities. See Ref.\cite{38} for the extension
of this method to a regime of weak field but with relativistic
velocities in the framework of a background-independent
Hamiltonian linearization of tetrad gravity.

\bigskip

Lacking solutions to Einstein's equations with matter
corresponding to simple systems to be used as idealizations for a
measuring apparatuses described by matter DO (hopefully also BO),
a generally covariant theory of measurement as yet does not exist.
We hope, however, that some of the clarifications achieved in this
paper of the existing ambiguities about observables will help in
developing such a theory.\bigskip

In the meanwhile we want to sketch here a scheme for implementing
- at least in principle - the physical individuation of points as
an experimental setup and protocol for positioning and
orientation. Our construction should be viewed in parallel to the
axiomatic treatment of Ehlers, Pirani and Schild. We could
reproduce the logical scheme of this axiomatic approach in the
following way.
\medskip

a) A {\it radar-gauge} system of coordinates can be defined in a
finite four-dimensional volume by means of a network of artificial
spacecrafts similar to the Global Position System (GPS) \cite{39}.
Let us consider a family of spacecrafts, whose navigation is
controlled from the Earth by means of the standard GPS. Note that
the GPS receivers are able to determine their actual position and
velocity because the GPS system is based on the advanced knowledge
of the gravitational field of the Earth and of the satellites'
trajectories, which in turn allows the {\it coordinate}
synchronization of the satellite clocks . During the navigation
the spacecrafts are test objects. Since the geometry of space-time
and the motion of the spacecrafts are not known in advance in our
case, we must think of the receivers as obtaining four, so to
speak, {\it conventional} coordinates by operating a full-ranging
protocol involving bi-directional communication to four {\it
super-GPS} that broadcast the time of their standard
a-synchronized clocks (see the discussion given in Ref.\cite{5}
and Refs.\cite{40} for other proposals in the same perspective).
This first step parallels the axiomatic construction of the {\it
conformal structure} of space-time.

Once the spacecrafts have arrived in regions with non weak fields,
like near the Sun or Jupiter, they become the (non test but with
world-lines assumed known from GPS space navigation) elements of
an experimental setup and protocol for the determination of a
local 4-coordinate system and of the associated 4-metric.

Each spacecraft, endowed with an atomic clock and a system of
gyroscopes, may be thought as a time-like observer (the spacecraft
world-line assumed known) with a tetrad (the time-like vector is
the spacecraft 4-velocity (assumed known) and the spatial triad is
built with gyroscopes) and one of them is chosen as the origin of
the radar-4-coordinates we want to define. This means that the
natural framework should be tetrad gravity instead of metric
gravity.

\medskip

b) At this point we have to synchronize the atomic clocks by means
of radar signals \cite{41}. As shown in I, in an Einstein
space-time there is a {\it dynamical determination of the
simultaneity convention}. However, since - again - the geometry of
space-time is not known in advance in our case, we could only lay
down the lines of an approximation procedure starting from an
arbitrary simultaneity convention like in special relativity. As
shown in Section VI of Ref.\cite{31},  the spacecraft $A$ chosen
as origin (and using the proper time $\tau$ along the assumed
known world-line) sends radar signals to the other spacecrafts,
where they are reflected back to $A$. For each radar signal sent
to a spacecraft $B$, the spacecraft $A$ records four data: the
emission time $\tau_o$, the emission angles $\theta_o$, $\phi_o$
and the absorption time $\tau_f$. Given four admissible (see
Ref.\cite{31}) functions ${\cal E}(\tau_o, \theta_o, \phi_o,
\tau_f)$, ${\vec {\cal G}}(\tau_o, \theta_o, \phi_o, \tau_f)$ the
point $P_B$ of the world-line of the spacecraft $B$, where the
signal is reflected, is given radar coordinates $\tau_{(R)}(P_B) =
\tau_o + {\cal E}(\tau_o, \theta_o, \phi_o, \tau_f)\, (\tau_f -
\tau_o)$, ${\vec \sigma}_{(R)}(P_B) = {\vec {\cal G}}(\tau_o,
\theta_o, \phi_o, \tau_f)$ and will be simultaneous (according to
this convention) to the point $Q$ on the world-line of the
spacecraft $A$ identified by $\tau {|}_Q = \tau_{(R)}(P_B)$.

This allows establishing a {\it radar-gauge system of
4-coordinates} (more exactly a coordinate grid) lacking any direct
metric content

\beq
 \sigma^A_{(R)} = ( \tau_{(R)}; \sigma^r_{(R)}),
 \label{VII1}
 \eeq

\noindent in a finite region, with $\tau_{(R)} = const$ defining
the radar simultaneity surfaces of this convention. By varying the
functions ${\cal E}$, ${\vec {\cal G}}$ we change the simultaneity
convention among the admissible ones \footnote{Einstein's
simultaneity convention corresponds to ${\cal E} = {1\over 2}$ and
to space-like hyper-planes as simultaneity surfaces.}.

\medskip

Note that by replacing {\it test radar signals} (conformal
structure) with {\it test particles} (projective structure) in the
measurements, we would define a different 4-coordinate system.
\medskip

Then the navigation system provides determination of the
4-velocities (time-like tetrads) of the satellites and the
${}^4g_{(R)\tau\tau}(\sigma_{(R)}^A)$ component of the 4-metric in
these coordinates. \medskip

In the framework of metric gravity the spacecrafts make repeated
measurements of the motion of {\it four} test particles. In this
way they test also the {\it projective structure} in a region of
space-time with a vacuum gravitational field. By the motion of
gyroscopes they measure the {\it shift} components ${}^4g_{(R)\tau
r}(\sigma^A_{(R)})$ of the 4-metric and end up (in principle) with
the determination of all the {\it components of the four-metric}
with respect to the {\it radar-gauge} coordinate system:

\beq
 {}^4g_{(R)AB}(\tau_{(R)}, \sigma^r_{(R)}).
  \label{VII2}
 \eeq

\medskip

The tetrad gravity alternative, employing test gyroscopes and
light signals (i.e. only the {\it conformal structure}), is the
following. By means of exchanges (one-way signals) of {\it
polarized} light it should be possible to determine how the
spatial triads of the satellites are rotated with respect to the
triad of the satellite chosen as origin (see also Ref.\cite{42}).
Once we have the tetrads ${}^4E^A_{(r)(\alpha )}(\tau_{(R)}, {\vec
\sigma}_{(R)})$ in radar coordinates, we can build from them the
inverse 4-metric ${}^4g_{(R)}^{AB}(\tau_{(R)}, {\vec
\sigma}_{(R)}) = {}^4E^A_{(r)(\alpha )}(\tau_{(R)}, {\vec
\sigma}_{(R)})\, {}^4\eta^{(\alpha )(\beta )}\, {}^4E^B_{(r)(\beta
)}(\tau_{(R)}, {\vec \sigma}_{(R)})$ in radar coordinates.

\medskip

c) By measuring the spatial and temporal variation of
${}^4{g}_{(R)AB}(\sigma^C_{(R)})$, the components of the Weyl
tensor and the Weyl eigenvalues can in principle be determined.
\bigskip

d) Points a), b) and c) furnishes {\it operationally} a slicing of
space-time into surfaces $\tau_{(R)} = const$, a system of
coordinates $\sigma^r_{(R)}$ on the surfaces, as well as a
determination of the components of the metric
${}^4{g}_{(R)AB}(\sigma^C_{(R)})$. The components of the Weyl
tensor (= Riemann in void) and the local value of the Weyl
eigenvalues, with respect to the radar-gauge coordinates
$(\tau_{(R)}, \sigma^r_{(R)})$ are also thereby determined. By
assuming the validity of Eintein's theory, it is then a matter of
computation:
\bigskip

i) To check whether Einstein's equation in radar-gauge coordinates
are satisfied. If not, this means that the chosen simultaneity
$\tau_{(R)} = const.$ is not the dynamical simultaneity of the
Einstein solution describing the solar system. By changing the
functions ${\cal E}$, ${\vec {\cal G}}$, we can put up an
approximation procedure converging towards the dynamical
simultaneity.\bigskip

ii) If $(\tau_R, {\vec \sigma}_R)$ are the radar coordinates
corresponding to the dynamical synchronization of clocks, we can
get a numerical determination of the intrinsic coordinate
functions ${\bar \sigma}^{\bar A}_R$ defining the radar gauge by
the gauge fixings $\sigma^A_R - {\bar \sigma}_R^{\bar A}(\sigma_R)
\approx 0$. Since we know the eigenvalues of the Weyl tensor in
the radar gauge, it is possible to solve in principle for the
functions $F^{\bar A}$ that reproduce the {\it radar-gauge}
coordinates as {\it radar-gauge intrinsic} coordinates

\beq \sigma^A_{(R)} = F^{\bar A}[{\tilde
\Lambda}^{(k)}_W[{}^3g(\tau ,\vec \sigma), {}^3\Pi(\tau ,\vec
\sigma)]],
 \label{VII3}
  \eeq

\bigskip

\noindent consistently with the {\it gauge-fixing} that enforces
just this particular system of coordinates:

\beq
 {\bar \chi}^A(\tau ,\vec \sigma ) {\buildrel {def} \over =}  \sigma^A -
 {\bar \sigma}^{\bar A}(\tau ,\vec \sigma ) = \sigma^A -
 F^{\bar A}\Big[{\tilde \Lambda}^{(k)}_W[{}^3g(\tau ,\vec \sigma), {}^3\Pi(\tau ,\vec
\sigma)]\Big] \approx 0.
 \label{VII4}
 \eeq

\noindent Finally, the {\it intrinsic coordinates} are
reconstructed as functions of the {\it DO} of the radar gauge, at
each point-event of space-time, as the identity

\beq \sigma^A \equiv {\bar \sigma}^{\bar A} = {\tilde F}^{\bar
A}_{G}[ r^{(R)}_{\bar a}(\tau ,\vec \sigma ), \pi^{(R)}_{\bar
a}(\tau , \vec \sigma)],
 \label{VII5}
 \eeq

\bigskip

This procedure of principle would close the {\it coordinative
circuit} of general relativity, linking individuation to
operational procedures \cite{5}.

\vfill\eject

\section{Bergmann Observables as Tensorial Dirac Observables and the Issue of
the Objectivity of Change.}

This Section is devoted to some crucial aspects of the definition
of {\it observable} in general relativity. While, for instance in
astrophysics, {\it matter observables} are usually defined as {\it
tetradic quantities} evaluated with respect to the tetrads of a
time-like observer so that they are obviously invariant under
${}_PDiff\, M^4$ (PDIQ), the definition of the notion of
observable for the gravitational field without matter faces a
dilemma. Two fundamental definitions of {\it observable} have been
proposed in the literature.
\bigskip

1) The {\it off-shell and on-shell Hamiltonian non-local Dirac
observables} (DO) \footnote{ For other approaches to the
observables of general relativity see Refs.\cite{43}: the {\it
perennials} introduced in this Reference are essentially our DO.
See Ref.\cite{44} for the difficulties in observing {\it
perennials} experimentally at the classical and quantum levels as
well as for their quantization. See Ref.\cite{45} about the
non-existence of observables built as spatial integrals of local
functions of Cauchy data and their first derivatives, in the case
of vacuum gravitational field in a closed universe. Also,
Rovelli's evolving constants of motion and partial observables
\cite{46} are related with DO; however, the holonomy loops used in
loop quantum gravity \cite{47} are PDIQ but not DO. On the other
hand, even recently Ashtekar \cite{48} noted that {\it The issue
of diffeomorphism invariant observables and practical methods of
computing their properties} is one among the relevant challenges.}
which, by construction, satisfy hyperbolic Hamilton equations of
motion and are, therefore, deterministically {\it predictable}. In
general, as already said, they are neither tensorial quantities
nor invariant under ${}_PDiff\, M^4$ (PDIQ).
\bigskip

2) The {\it configurational Bergmann observables} ({\it BO})
\cite{22}: they are quantities defined on $M^4$ {\it which not
only are independent of the choice of the coordinates}, [i.e. they
are either scalars or invariants\footnote{In Ref.\cite{19}
Bergmann defines: i) a {\it scalar} as a local field variable
which retains its numerical value at the same world point under
coordinate transformations (passive diffeomorphisms),
$\varphi^{'}(x^{'}) = \varphi (x)$; ii) an {\it invariant} $I$ as
a functional of the given fields which has been constructed so
that if we substitute the coordinate transforms of the field
variables into the argument of $I$ instead of the originally given
field variables, then the numerical value of $I$ remains
unchanged. } under ${}_PDiff\, M^4$ (PDIQ)], but are also "{\it
uniquely predictable from the initial data}". An equivalent, but
according to Bergmann more useful, definition of a (PIDQ) BO, is
"{\it a quantity that is invariant under a coordinate
transformation that leaves the initial data unchanged}".

\medskip

Let us note, first of all, that PDIQ's are {\it not} in general
DO, because they may also depend on the eight gauge variables $n$,
$n_r$, $\xi^r$, $\pi_{\phi}$. Thus most, if not all, of the
curvature scalars are gauge dependent quantities at least at the
kinematic off-shell level. For example, each 3-metric in the
conformal gauge orbit has a different 3-Riemann tensor and
different 3-curvature scalars. Since 4-tensors and 4-curvature
scalars depend: i) on the lapse and shift functions (and their
gradients); ii) on $\pi_{\phi}$, both implicitly and explicitly
through the solution of the Lichnerowicz equation (and this
affects the 3-curvature scalars), most of these objects are in
general gauge dependent variables from the Hamiltonian point of
view. The simplest relevant off-shell scalars with respect to
${}_PDiff\, M^4$, which exhibit such gauge dependence, are the
bilinears ${}^4R_{\mu\nu\rho\sigma}\, {}^4R^{\mu\nu\rho\sigma}$,
${}^4R _{\mu\nu\rho\sigma}\, \epsilon^{\mu\nu\alpha\beta}\,
{}^4R_{\alpha\beta}{} ^{\rho\sigma}$  and the {\it four
eigenvalues of the Weyl tensor} exploited in Section V. What said
here {\it does hold, in particular, for the line element $ds^{2}$
and, therefore, for the causal structure of space-time}.
\medskip

On the other hand, BO are those special PDIQ which are also {\it
predictable}. Yet, the crucial question is now "{\it what does it
precisely mean to be predictable within the configurational
framework} ?". Bergmann, gave in fact a third definition of BO or,
better, a {\it third part} of the original definition, as "a
dynamical variable that (from the Hamiltonian point of view) has
vanishing Poisson brackets with all the constraints", i.e.,
essentially, is also a DO. This means that Bergmann thought,
though only implicitly and without proof, that {\it
predictability} implied that a BO must {\it also} be {\it
projectable to phase space to a special subset of DO that are also
PDIQ}.

\bigskip

The unresolved multiplicity of Bergmann's definitions leads to an
entangled net of problems. First of all, as shown at length in
Ref.\cite{3}, in order to tackle the Cauchy problem at the
configuration level \footnote{ In the theory of systems of partial
differential equations this is done in a {\it passive} way in a
given coordinate system and then extended to all coordinate
systems.} one has firstly to disentangle the Lagrangian
constraints from Einstein's equations, then to take into account
the Bianchi identities, and finally to write down a system of
hyperbolic equations. As a matter of fact one has to mimic the
Hamiltonian approach, but with the additional burden of lacking an
algorithm for selecting those predictable configurational field
variables whose Hamiltonian counterparts are just the DO. The only
thing one might do is to adopt an inverse Legendre transformation,
to be performed after the Shanmugadhasan canonical transformation
characterizing a possible complete set of DO. Yet, this just
corresponds to the inverse of Bergmann's statement that the BO are
projectable to special (PDIQ) DO. In conclusion, {\it
configurational predictability must be equivalent to the statement
of off-shell Hamiltonian gauge invariance}. The moral is that the
complexity of the issue should warn against any naive utilization
of geometric intuitions in dealing with the initial value problem
of general relativity within the configurational approach.
\medskip

This Hamiltonian predictability of BO entails in turn that only
{\it four functionally independent BO can exist for the vacuum
gravitational field}, since the latter has only two pairs of
conjugate independent degrees of freedom. Let us see now why
Bergmann's multiple definition of BO raises additional subtle
problems.
\medskip

Bergmann himself proposed a constructive procedure for the BO.
This is essentially based on his re-interpretation of Einstein's
{\it coincidence argument} in terms of the individuation of
space-time points as point-events by using {\it intrinsic
pseudo-coordinates}. In his - already quoted - words \cite{22}:
"we might then identify a {\it world point}
(location-plus-instant-in-time) by the values assumed by (the four
intrinsic pseudo-coordinates) and ask for the value, there and
then, of a fifth field". As an instantiation of this procedure,
Bergmann refers to Komar's \cite{20} pseudo-tensorial
transformation of the 4-metric tensor to the {\it intrinsic
pseudo-coordinate} system [$\sigma^A = \sigma^A ({\bar
\sigma}^{\bar A})$ is the inversion of Eqs.(\ref{IV2})]

\beq
 {}^4{\bar g}_{\bar A\bar B}({\bar \sigma}^{\bar C} ) = {{\partial \sigma^C}\over
{\partial {\bar \sigma}^{\bar A} }}\, {{\partial \sigma^D}\over
{\partial {\bar \sigma}^{\bar B}}}\, {}^4g_{CD}(\sigma ).
 \label{V1}
 \eeq

\noindent The ${}^4{\bar g}_{\bar A\bar B}$ represent {\it ten
invariant scalar} \,(PDIQ or {\it tetradic}) {\it components} of
the metric; of course, they are not all independent since must
satisfy eight functional restrictions following from Einstein's
equations.\medskip

Now, Bergmann {\it claims} that the ten components $ {}^4{\bar
g}_{\bar A\bar B}({\bar \sigma}^{\bar C} )$ are a complete, but
non-minimal, set of BO. {\it This claim, however, cannot be true}.
As already pointed out, since BO are predictable they must in fact
be equivalent to (PDIQ) DO so that, for the vacuum gravitational
field, {\it exactly four} functions at most, out of the ten
components ${}^4{\bar g}_{\bar A\bar B}({\bar \sigma}^{\bar C} )$,
can be simultaneously BO and DO, while the remaining components
must be {\it non-predictable} PDIQ, counterparts of ordinary
Hamiltonian gauge variables.\medskip

On the other hand, as shown in Section II, the four independent
degrees of freedom of the pure vacuum gravitational field, even
for Bergmann, {\it have allegedly already been exploited for the
individuation of point-events}. Besides, as Bergmann explicitly
asserts in his purely {\it passive} interpretation, the PDIQ  $
{}^4{\bar g}_{\bar A\bar B}({\bar \sigma}^{\bar C} )$ identify
{\it on-shell} a 4-geometry, i.e. an equivalence class in
${}^4Geom = {}^4Riem / {}_PDiff\, M^4$. Furthermore, as shown in
Section II of I, Eq.(2.8), the identification of the same
4-geometry starting from active diffeomorphisms can be done by
using their passive re-formulation (the group $Q$). Finally, let
us remark that Bergmann's intention to first exploit the intrinsic
pseudo-coordinates and then "ask for the value, there and then, of
a fifth field" makes sense only if such "{\it fifth field}" is a
{\it matter field}. Asking the question for purely gravitational
quantities like ${}^4{\bar g}_{\bar A\bar B}({\bar \sigma}^{\bar
C} )$ would be at least tautological since, as we have seen, only
four of them can be independent and have already been exploited.
If the individuation procedure is intended to be effective, it
would make little sense to assert that point-events have such and
such values in terms of point-events.

\medskip

But now, Bergmann's incorrect claim is relevant also to another
interesting quandary. Indeed, Bergmann's {\it main}
configurational notion of {\it observable} and its implications
are accepted as they stand in a paper of John Earmann,
Ref.\cite{25}. In particular Earman notes that the intrinsic
coordinates [${\bar \sigma}^{\bar C}$] can be used to support
Bergmann's observables and says "one can speak of the event of the
metric - components - $ {}^4{\bar g}_{\bar A\bar B}({\bar
\sigma}^{\bar C} )$ - having - such - and - such - values - in -
the - coordinate - system - $\{ {\bar \sigma}^{\bar C}  \}$ - at -
the - location - where - the - ${\bar \sigma}^{\bar C}$ - take -
on - values - such - and - so" and (aptly) calls such an item a
{\it Komar event}, adding moreover that "the fact that a given
Komar event occurs (or fails to occur) is an observable matter in
Bergmann's sense, albeit in an abstract sense because how the
occurrence of a Komar event is to be observed/measured is an
unresolved issue". Earman's principal aim, however, is to exploit
Bergmann's definition of BO to show that "it implies that there is
no physical change, i.e., no change in the observable quantities,
at least not for those quantities that are constructible in the
most straightforward way from the materials at hand". Although we
are not committed here to object to what Earman calls "modern
Mc-Taggart argument" about change, we are obliged to take issue
against Earman's radical {\it universal no-change argument}
because, if sound, it would contradict the substance of Bergmann's
definition of predictability and would jeopardize the relation
between BO and DO which is fundamental to our program.\medskip

In order to scrutinize this point, let us resume, for the sake of
clarity, the essential basic ingredients of the present
discussion.

{\it One}: the equations of motion derived from Einstein-Hilbert
Action and those derived from ADM Action have {\it exactly the
same physical content}: the ADM Lagrangian leads, through the
Legendre transformation, to Hamilton equations equivalent to
Einstein's equations.

{\it Two}: Hamiltonian predictability must, therefore, be
equivalent to Lagrangian predictability: specification of the
latter, however, is an awkward task.

{\it Three}: the only functionally independent Hamiltonian
predictable quantities for the vacuum gravitational field, are
four DO.

{\it Four}: by inverse Legendre transformation, every DO has {\it
a Lagrangian predictable counterpart}.
\medskip

Then, a priori, one among the following three possibilities might
be true: i) all the existing BO must also be DO; this means
however that only four functionally independent BO can exist; ii)
some of the existing BO are also DO while other are not; iii) no
one of the existing BO is also a DO. Possibilities ii) and iii)
entail that Bergmann's multiple definition (that including the
third part) of BO is inconsistent, so that no BO satisfying such
multiple definition would exist. Yet, the third part of Bergmann
definition is essential for the overall meaning of it since no
Lagrangian definition of predictability {\it independent of its
Hamiltonian counterpart} can exist because of {\it Two}. Thus
cases ii) and iii) imply inconsistency of the very concept of
Bergmann's observability. Of course, it could be that even i) is
false since, after all, Bergmann did not prove the
self-consistency of his multiple definition: but this would mean
that no Lagrangian predictable quantity could exist which
simultaneously be a PDIQ. Here, we are {\it assuming} that
Bergmann's multiple definition is consistent and that i) is true.
We will formalize this assumption into a definite {\it
constructive conjecture} later on in this Section.
\bigskip

Let us take up again the discussion about the reality of change.
As already noted, the discussion in terms of BO in the language of
{\it Komar events} (or {\it coincidences}) must be restricted to
the properties of matter fields because, consistently with the
multiple Bergmann's definition, only four of the BO can be purely
gravitational in nature. And, if these latter have already been
exploited for the individuation procedure, it would again make
little sense to ask whether point-events do or do not change.
Therefore, let us consider Earman's argument by examining his
interpretation of predictability and the consequent implications
for a BO, say $B(p)$, $p \in M^4$, which, besides depending on the
4-metric and its derivatives up to some finite order, also depends
on matter variables, and is of course a PDIQ. In order to simplify
the argument, Earman concentrates on the special case of the
vacuum solutions to the Einstein's field equations, asserting
however that the argument easily generalizes to non-vacuum
solutions. Since we have already excluded the case of vacuum
solutions, let us take for granted that this generalization is
sound. Earman argues essentially in the following way: 1) There
are existence and uniqueness proofs for the initial value problem
of Einstein's equations, which show that for appropriate initial
data associated to a three manifold $\Sigma_o \subset M^4$, there
is a unique {\it up to diffeomorphism} (obviously to be intended
{\it active}) \footnote{ Note that Earman deliberately deviates
here from the purely {\it passive viewpoint} of Bergmann (and of
the standard Cauchy problem for partial differential equations) by
resorting to {\it active} diffeomorphisms in place of the {\it
coordinate transformations that leave the initial data unchanged}
or, possibly, in place of their extension in terms of the passive
re-interpretation of active diffeomorphisms ($Q$ group).} maximal
development for which $\Sigma_o$ is a Cauchy surface; 2) By
definition, a BO is a PDIQ whose value $B(p)$ at some point $p$ in
the future of $\Sigma_o$ is predictable from initial data on
$\Sigma_o$. If $D_A : M^4 \mapsto M^4$ is an {\it active}
diffeomorphism that leaves $\Sigma_o$ and its past fixed, the
point $p$ will be sent to the point $p^{'} = D_A \cdot p$. Then,
the general covariance of Einstein's equations, {\it conjoined
with predictability}, is interpreted to imply $B(p) = B(D_A \cdot
p)$. This result, together with the definition $B'(p) =
B(D_A^{-1}\cdot p)$ of the drag along of $B$ under the active
diffeomorphism $D_A$, entails $D_A^*B = B$ for a BO. In
conclusion, since $\Sigma_o$ is arbitrary, a matter BO should be
{\it constant everywhere} in $M^4$.

\bigskip

It is clear that, within our class of space-times, this conclusion
cannot hold true for any matter dependent BO that is projectable
to the DO of the gravitational field {\it cum} matter, if only for
the fact that such BO are in fact ruled by the weak {\it ADM
energy} which generates real temporal change (see Section IIID,
Eq.(3.8), of I).\, The crucial point in Earman's argument is the
assertion that predictability implies $B(p) = B(\phi \cdot p)$.
But this does {\it not} correspond to the property of {\it
off-shell gauge invariance} spelled above as the main
qualification of predictable quantities, except of course for the
trivial case of quantities everywhere constant. As clarified in
Sections II and III of I, the relations between active
diffeomorphisms and gauge transformation (which are necessarily
involved by the DO) is not straightforward. Precisely, because of
the properties of the group $Q$ of Bergmann and Komar, we have to
distinguish between the {\it active} diffeomorphisms in Q that
{\it do belong} to $Q_{can}$ and those that {\it do not belong} to
$Q_{can}$. Actually recall that:

i) The intersection $Q_{can} \cap {}_PDiff\, M^4$ identifies the
space-time {\it passive} diffeomorphisms which, respecting the 3+1
splitting of space-time, are {\it projectable} to ${\cal G}_{4\,
P}$ in phase space;

ii) The remaining elements of $Q_{can}$ are the {\it projectable}
subset of {\it active diffeomorphisms} in their passive view. The
union $Q_{can} \cup {}_PDiff\, M^4$ exhausts the Hamiltonian view
of Leibniz equivalence.

iii) The elements of Q which do not belong to $Q_{can}$ are not
projectable to phase space at all and have, therefore, nothing to
do with Lagrangian {\it predictability}. In particular the
non-projectable active diffeomorphisms (passively reinterpreted)
do not correspond to Hamiltonian gauge transformations acting
within a given universe, solution of Einstein's equations.
Actually many of them are maps on the space of Cauchy data (i.e.
maps among different universes) and consequently are unrelated to
Leibniz equivalence.

\medskip

In conclusion, for most {\it active} diffeomorphisms\footnote{ In
particular, the special $D_A$'s considered by Earman.}, the
conclusion $B(p) = B(D_A \cdot p)$ cannot hold true. This
erroneous conclusion seems to b e just an instantiation of how
misleading may be any loose geometrical and non-algorithmic
interpretation of $\Sigma_o$ as a Cauchy surface within the
Lagrangian configuration approach to the initial value problem of
general relativity.

\bigskip

Having settled this important point, let us come back to tetradic
fields. Besides the tetradic components (\ref{V1}) of the
4-metric, we have to take into account the extrinsic curvature
tensor ${}^3K^{AB}(\sigma ) = {{\partial \sigma^A}\over {\partial
x^{\mu}}}\, {{\partial \sigma^B}\over {\partial x^{\nu}}}\,
{}^3K^{\mu\nu}(x)$. In the coordinates $\sigma^A$ adapted to
$\Sigma_{\tau}$, it has the components ${}^3K^{\tau\tau}(\sigma )
= {}^3K^{\tau r}(\sigma ) = 0$ and ${}^3K^{rs}(\sigma )$ and we
can rewrite it as

\beq
 {}^3{\bar K}^{\bar A\bar B}({\bar \sigma}^{\bar C}) = {{\partial
 {\bar \sigma}^{\bar A}}\over {\partial \sigma^A}}\, {{\partial
 {\bar \sigma}^{\bar B}}\over {\partial \sigma^B}}\, {}^3K^{AB}(\sigma )
 =  {{\partial
 {\bar \sigma}^{\bar A}}\over {\partial \sigma^r}}\, {{\partial
 {\bar \sigma}^{\bar B}}\over {\partial \sigma^s}}\, {}^3K^{rs}(\sigma ).
 \label{V2}
 \eeq

\noindent In this way we get 10 additional scalar (tetradic)
quantities (only six of which are independent due to the vanishing
of the lapse and shift momenta) replacing ${}^3K^{rs}(\sigma )$
and, therefore, the ADM momenta ${}^3{\tilde \Pi}^{rs}(\sigma ) =
\epsilon k\, [\sqrt{\gamma}\, ({}^3K^{rs} - {}^3g^{rs}\, {}^3K
)](\sigma )$.

In each intrinsic coordinate system ${\bar \sigma}^{\bar A} =
F^{\bar A}[\Lambda^{(k)}_W(\sigma )]$, we have consequently the 20
scalar (tetradic) components ${}^3{\bar g}_{\bar A\bar B}({\bar
\sigma}^{\bar C})$ and ${}^3{\bar K}^{\bar A\bar B}({\bar
\sigma}^{\bar C})$ of Eqs.(\ref{V1}), (\ref{V2}), only 16 of which
are functionally independent. However, four of them are scalar
intrinsic constraints  ${\bar {\cal H}}^{\bar A}({\bar
\sigma}^{\bar C}) = {{\partial {\bar \sigma}^{\bar A}}\over
{\partial \sigma^A}}\, {\cal H}^A(\sigma ) \approx 0$ replacing
the super-hamiltonian and super-momentum constraints ${\cal
H}^a(\sigma ) = \Big( {\cal H}(\sigma ); {\cal H}^r(\sigma ) \Big)
\approx 0$.

\bigskip

The various aspects of the discussion given above strongly suggest
that, in order to give consistency to Bergmann's unresolved
multiple definition of BO and, in particular, to his (strictly
speaking unproven) claim \cite{22} about the {\it existence} of DO
that are simultaneously (PDIQ) BO, the following {\it conjecture}
should be true:

\bigskip
\noindent {\bf A Main Conjecture}: "The Darboux basis whose 16 ADM
variables consist of the 8 Hamiltonian gauge variables $n$, $n_r$,
$\xi^r$, $\pi_{\phi}$, the 3 Abelianized constraints ${\tilde
\pi}_r^{{\vec {\cal H}}} \approx 0$, the conformal factor $\phi$
(to be determined by the super-hamiltonian constraint) and the
(non-tensorial) DO $r_{\bar a}$, $\pi_{\bar a}$, appearing in the
quasi-Shanmugadhasan canonical basis (3.7) of I {\it can be
replaced} by a Darboux basis whose 16 variables are all PDIQ (or
tetradic variables), such that four of them are simultaneous DO
and BO, eight vanish because of the first class constraints, and
the other 8 are coordinate-independent gauge variables.\footnote{
Note that Bergmann's constructive method based on tetradic
4-metric is not by itself conclusive in this respect !}"
\bigskip

If this conjecture is sound, it would be possible to construct an
{\it intrinsic} Darboux basis  of the Shanmugadhasan type
(Eq.(3.7) of I). Then a suitable transformation performed
off-shell before adding the gauge fixings $\sigma^A - {\bar
\sigma}^{\bar A}(\sigma ) \approx 0$, should exist bringing from
the non-tensorial Darboux basis (3.7) of I to this new {\it
intrinsic} basis. Since the final result would be a representation
of the gauge variables as coordinate-independent (PIDQ) gauge
variables and of the DO as {\it Dirac-and-Bergmann observables},
the freedom of the above transformation reduces to the possibility
of mixing the PDIQ gauge variables among themselves and of making
canonical transformations in the subspace of the Dirac-Bergmann
observables.\medskip

More precisely, we would have a family of quasi-Shanmugadhasan
canonical bases in which all the variables are PDIQ and include 7
PDIQ first class constraints (not the one corresponding to the
super-hamiltonian constraint) that play the role of momenta. It
would be interesting, in particular, to check the form of the
constraint replacing the standard super-hamiltonian constraint. By
re-expressing the 4 Weyl eigenvalues in terms of anyone of these
PDIQ canonical bases, we could still define a Hamiltonian gauge,
namely an on-shell 4-coordinate system and then derive the
associated individuation of point-events by means of gauge-fixings
of the type (\ref{IV5}). {\it Note that this would break general
covariance even if the canonical basis is PDIQ} ! The only
difference with respect to the standard Hamiltonian bases would be
that, instead of being non-tensorial quantities, both $r^G_a$,
$\pi^G_a$ and ${\tilde F}^{\bar A}_G$ in Eq.(\ref{IV8}) would be
PDIQ.

\medskip

As anticipated in the Introduction, further strong support to the
conjecture comes from Newman-Penrose formalism \cite{23} where the
basic tetradic fields are the 20 Weyl and Ricci scalars which are
PDIQ by construction . While the vanishing of the Ricci scalars is
equivalent to Einstein's equations (and therefore to a scalar form
of the super-hamiltonian and super-momentum constraints), the 10
Weyl scalars plus 10 scalars describing the ADM momenta
(restricted by the four primary constraints) should lead to the
construction of a Darboux basis spanned only by PDIQ restricted by
eight PDIQ first class constraints. Again, a quasi-Shanmugadhasan
transformation should produce the Darboux basis of the conjecture.
The problem of the phase space re-formulation of Newman-Penrose
formalism is now under investigation.
\bigskip

A final important logical component of the issue of the
objectivity of change is the particular question of {\it temporal}
change. This aspect of the issue is not usually tackled as a
sub-case of Earman's {\it no-universal-change} argument discussed
above in terms of BO, so it should be answered separately. We
shall confine our remarks to the objections raised by Belot and
Earmann \cite{49} and Earman \cite{25} (see also
Refs.\cite{49,50,51,52} for the so called {\it problem of time} in
general). According to these authors, the reduced phase space of
general relativity is a frozen space without evolution. Belot and
Earman draw far reaching conclusions about the absence of real
(temporal) change in general relativity from the circumstance
that, in {\it spatially compact} models of general relativity, the
Hamiltonian temporal evolution boils down to a mere gauge
transformation and is, therefore, physically meaningless. We want
to stress, however, that this result {\it does not apply to all
families of Einstein space-times}. In particular, there exist
space-times like the Christodoulou-Klainermann space-times
\cite{4} we are using in this paper that constitute a {\it
counterexample to the frozen time argument}. They are defined by
suitable boundary conditions, are globally hyperbolic,
non-compact, and asymptotically flat at spatial infinity as shown
in Section III of I. The existence of such meaningful
counterexamples entails, of course, that we are not allowed to
draw negative conclusions in general about the issue of temporal
change in general relativity.

\medskip

We can conclude that in these space-times there is neither a
frozen reduced phase space nor a Wheeler-DeWitt interpretation
based on some local concept of time like in compact space-times.
Therefore, {\it our models of general relativity are perfectly
adequate to accommodate objective temporal change}.

\bigskip

Let us remark that the definitions given in this Section of the
{\it notion of observable} in general relativity are in
correspondence with two different points of view, existing in the
physical literature, that are clearly spelled out in Ref.\cite{53}
and related references, namely:\hfill\break

i) The {\it non-local point of view} of Dirac \cite{54}, according
to which causality implies that only gauge-invariant quantities,
i.e., DO, can be measured. As we have shown, this point of view is
consistent with general covariance. For instance, ${}^4R(\tau
,\vec \sigma )$ is a scalar under diffeomorphisms, and therefore a
BO,  but it is not a DO - at least the kinematic level - and
therefore, according to Dirac, not an observable quantity. Even if
${}^4R(\tau ,\vec \sigma )\, {\buildrel \circ \over =}\, 0$ in
absence of matter, the other curvature scalars do not vanish in
force of Einstein's equations and, lacking known solutions without
Killing vectors, it is not clear  their connection with the DO.
The 4-metric tensor ${}^4g_{\mu\nu}$ itself as well as the line
element $ds^2$ are not DO so a completely fixed gauge is needed to
get a definite functional form for them in terms of the DO in that
gauge. This means that all standard procedures for defining
measures of length and time \cite{55,56,57} and the very
definition of angle and distance properties of the material bodies
forming the reference system, are {\it gauge dependent}. Then they
are determined only after a complete gauge fixing and after the
restriction to the solutions of Einstein's equations has been
made\footnote{Note that in standard textbooks these procedures are
always defined without any reference to Einstein's equations.}.
Likewise, only after a complete gauge fixing the procedure for
measuring the Riemann tensor with $n \geq 5$ test particles
described in Ref.\cite{57} (see also Ref.\cite{21}) becomes
completely meaningful, just as it happens for the electro-magnetic
vector potential in the radiation gauge.

Note finally that, after the introduction of matter, even the
measuring apparatuses should be described by the gauge invariant
matter DO associated with the given gauge. \hfill\break

ii) The {\it local point of view}, according to which the
space-time manifold $M^4$ is a kind of postulated (often without
any explicit statement) background manifold of physically
determinate {\it events}, like it happens in special relativity
with its absolute chrono-geometric structure. Space-time points
are assumed physically distinguishable, because any measurement is
performed in the frame of a given reference system interpreted as
a physical laboratory. In this view the gauge freedom of generally
covariant theories is reduced to mere passive coordinate
transformations. See for instance Ref. \cite{58} for a refusal of
the concept of DO in general relativity based on the local point
of view. This point of view, however, discount the Hole Argument
completely and must renounce to a deterministic evolution, so that
it is ruled out by our results.
\bigskip

Rovelli (\cite{53}) accepts the non-local point of view and
proposes to introduce some special kind of matter for defining a
{\it material reference system} (not to be confused with a
coordinate system) to localize points in $M^4$. The aim is to
recover the local point of view in some approximate way
\footnote{The main approximations are: 1) neglect, in Einstein
equations, the energy-momentum tensor of the matter forming the
material reference system (it's similar to what happens for test
particles); 2) neglect, in the system of dynamical equations, the
entire set of equations determining the motion of the matter of
the reference system (this introduces some {\it indeterminism} in
the evolution of the entire system).} since the analysis of
classical experiments shows that both approaches lead to the same
conclusions in the weak field regime. This approach relies
therefore upon {\it matter} to solve the problem of the {\it
individuation} of space-time points as point-events, at the
expense of loosing determinism. The emphasis on the fundamental
role of matter for the individuation issue is present also in
Refs.\cite{36,59,52}, where {\it material clocks} and {\it
reference fluids} are exploited as test matter.

As we have shown, however, the problem of the individuation can be
solved {\it before and without} the introduction of matter. The
presence of matter has the only effect of {\it modifying} the
individuation and, of course, is fundamental in trying to
establish a general-relativistic theory of measurement.

\vfill\eject

\section{Concluding Remarks.}

\noindent The aim of this paper and the previous one (I) was to
show that the Hamiltonian approach to general relativity in the
ADM formulations has the capability to get new insights into both
deep foundational issues and technical problems of the theory,
including its experimental forefront.
\medskip

In paper I we have clarified  the correspondence between the {\it
active diffeomorphisms} operating in the configurational manifold
$M^4$, on the one hand, and the {\it on-shell gauge
transformations} of the ADM canonical approach to general
relativity, on the other. Understanding such correspondence is
fundamental for fully disclosing the connection of the Hole
phenomenology, at the Lagrangian level, with the correct
formulation of the initial value problem of the theory and its
gauge invariance, at the Hamiltonian level. The upshot is the
discovery that, as concerns both the Hole Argument and the issue
of {\it predictability}, only the active diffeomorphisms of $M^4$,
which are also elements of $Q_{can}$ (i.e. only the projectable
maps of Ref.\cite{10}), play an effective role to define a correct
mathematical setting of the initial value problem at the
Lagrangian level.

\medskip

Secondly, we have identified a class of solutions of Einstein's
equations (of the type of the Christodoulou-Klainermann
space-times \cite{7}), which are particularly interesting for both
our main program and a unified description of gravity and particle
physics as well as the analysis of the gravitational phenomenology
in the solar system. Such class allows in particular: i)
exploiting the 3+1 splitting of space-time required by the ADM
Hamiltonian approach to general relativity; ii) an effective time
evolution ruled by the so-called weak ADM energy: they provide
thereby a counterexample to the frozen time argument and are free
of any Wheeler-De Witt interpretation; iii) a possible
accomodation of the standard model of elementary particles; iv)
the vanishing of super-translations and consequent definability of
the total angular momentum; v) the definition of asymptotic
idealized structures playing the role of the {\it fixed stars} of
the empirical astronomy. Finally, by means of suitable
restrictions upon the admissible simultaneity hyper-surfaces, they
become Lichnerowicz manifolds \cite{60} and allow thereby for the
existence of generalized Fourier transforms and the definition of
positive and negative asymptotic frequencies. The last option
paves the way for a quantization program.

\medskip

Within this background, we have shown that, unlike the case of
standard special relativity, the admissible notions of distant
simultaneity in canonical metric gravity turn out to be {\it
dynamically determined on-shell}, while off-shell  different {\it
conventions} within the same {\it universe} are merely different
{\it gauge-related options} like in special relativity \cite{31}.
This gives new insight into the old - and outdated - debate about
the so-called {\it conventionality} of distant simultaneity in
special relativity, showing the trading between {\it
conventionality} and {\it gauge freedom}. On this backdrop, we
have furthermore recognized the distinct physical role played by
the DO, as embodying {\it tidal-like} dynamical effects, on the
one hand, and that played by {\it off shell} gauge variables as
connected to {\it generalized inertial} effects, on the other.
\bigskip

The main results of the present paper are:
\medskip

\noindent 1) A definite procedure for the {\it physical
individuation} of the mathematical points of the would-be
space-time manifold $M^4$ into {\it physical  point-events},
through a gauge-fixing {\it identifying} the mathematical
4-coordinates with the intrinsic pseudo-coordinates of Komar and
Bergmann (defined as suitable functionals of the Weyl scalars).
This has led to the conclusion that each of the point-events of
space-time is endowed with its own physical individuation as the
value, as it were, at that point, of the four canonical
coordinates or DO (just four!), which describe the dynamical
degrees of freedom of the gravitational field. Since such degrees
of freedom are non-local functionals of the 3-metric and
3-curvature \footnote{Admittedly, at least at the classical level,
we don't know of any detailed analysis of the relationship between
the notion of {\it non-local observable} (the predictable degrees
of freedom of a gauge system), on one hand, and the notion of a
quantity which has to be operationally {\it measurable} by means
of {\it local apparatuses}, on the other. Note that this is true
even for the simple case of the electro-magnetic field where the
Dirac observables are defined by the {\it transverse} vector
potential and the {\it transverse} electric field. Knowledge of
such fields at a definite mathematical time involves data on the
whole Cauchy surface at that time. Even more complex is the
situation in the case of Yang-Mills theories \cite{61}}, they are
unsolvably entangled with the whole metrical texture of the
simultaneity surfaces in a way that is strongly both
gauge-dependent and highly non-local with respect to the
background mathematical coordinatization. Still, once they are
calculated, they appear as {\it local fields} in terms of the
background {\it mathematical} coordinatization, a fact that makes
the identity Eq.(\ref{IV8}) possible and shows, in a sense, a
Machian flavor within a non-Machian environment. We can also say,
on the other hand, that {\it any coordinatization} of the manifold
can be seen as embodying the physical individuation of points,
because it can be implemented as the Komar-Bergmann {\it intrinsic
pseudo-coordinates}, after a suitable choice of the functions of
the Weyl scalars and of the gauge-fixing. Moreover, as stressed in
Section III, {\it only on-shell matter will come to influence the
very physical individuation of points}. We claim that our results
bring the Synge-Bergmann-Komar-Stachel program of the physical
individuation of space-time points to its natural end.
\medskip

\medskip

\noindent 2) It should be clear by now that the Hole Argument has
little to do with an alleged \emph{indeterminism} of general
relativity as a dynamical theory. For, in our analysis of the
initial-value problem within the Hamiltonian framework, we have
shown that {\it on shell} a complete gauge-fixing (which could in
theory concern the whole space-time) is equivalent, {\it among
other things}, to the choice of an atlas of coordinate charts on
the space-time manifold, and in particular {\it within the Hole}.
At the same time, we have shown that a peculiar subset of the
active diffeomorphisms of the manifold can be interpreted as
passive Hamiltonian gauge transformations. Actually, only this
subset, realizing the essential content of Leibnitz equivalence,
plays an effective role in connection to the Hole Argument.

\medskip

\noindent 3) An outline of the implementation (in principle) of
the physical individuation of point-events as an experimental
setup and protocol for positioning and orientation, which closes,
as it were, the practical {\it coordinative circuit} of general
relativity.
\medskip

\noindent 4) A clarification of Bergmann's multiple ambiguous
definition of {\it observable} in general relativity. This has led
to formulate our {\it main conjecture} concerning the unification
of Bergmann's and Dirac's concepts of observable, as well as to
restate the issue of change and, in particular and independently,
of temporal change, within the Hamiltonian approach to Einstein
equations. When concretely carried out, this program would provide
even {\it explicitly} evidence for the {\it invariant objectivity}
of point-events. Furthermore, the existence of simultaneous {\it
Bergmann-Dirac observables} and {\it PDIQ gauge variables} would
lead to a description of tidal-like and inertial-like effects in a
coordinate independent way, while the {\it Dirac-Bergmann}
observables only would remain as the only quantities subjected to
a causal evolution. If the conjecture about the existence of
simultaneous DO-BO observables is sound, it would open the
possibility of a new type of coordinate-independent canonical
quantization of the gravitational field. Only the DO should be
quantized in this approach, while the gauge variables, i.e., the
{\it appearances} of inertial effects, should be treated as
c-number fields (a prototype of this quantization procedure is
under investigation \cite{62} in the case of special relativistic
and non-relativistic quantum mechanics in non-inertial frames).
This would permit to preserve causality (the space-like character
of the simultaneity Cauchy 3-surfaces), the property of having
only the 3-metric quantized (with implications similar to loop
quantum gravity for the quantization of spatial quantities), and
to avoid any talk of {\it quantization of the 4-geometry} (see
more below), a talk we believe to be deeply misleading (in this
connection see Ref. \cite{63})

\bigskip

We want to conclude our discussion with some general remarks about
the foundation of general relativity and some venture-some
suggestions concerning quantum gravity.

\bigskip

A) First of all, our program is substantially grounded upon the
{\it gauge nature} of general relativity. Such property of the
theory, however, is far from being a simple matter and we believe
that it is not usually spelled out in a sufficiently explicit and
clear fashion. The crucial point is that general relativity {\it
is not} a standard gauge theory like, e.g., electromagnetism or
Yang-Mills theories in some relevant respects. The relevant fact
is that, while from the point of view of the constrained
Hamiltonian formalism general relativity is a gauge theory like
any other, it is radically different from the physical point of
view. For, in addition to creating the distinction between what is
observable\footnote{In the Dirac or Bergmann sense.} and what is
not, the gauge freedom of general relativity is unavoidably
entangled with the definition--constitution of the very {\it
stage}, space--time, where the {\it play} of physics is enacted.
In other words, the gauge mechanism has the double role of making
the dynamics unique (as in all gauge theories), and of fixing the
spatio-temporal reference background. It is only after a complete
gauge-fixing is specified (i.e. after the individuation of a {\it
well defined} physical laboratory network that we have called a
{\it global non-inertial space-time laboratory}), and after going
on shell, that even the mathematical manifold $M^4$ gets a {\it
physical individuation} and becomes the spatio-temporal carrier of
well defined physical {\it tidal-like} and {\it generalized
inertial} effects.

In gauge theories such as electromagnetism, we can rely from the
beginning on empirically validated, gauge-invariant dynamical
equations for the {\it local} fields. This is not the case for
general relativity: in order to get dynamical equations for the
basic field in a {\it local} form, we must pay the price of
Einstein's general covariance which, by ruling out any background
structure at the outset, weakens the objectivity that the
spatiotemporal description could have had {\it a priori}.

\medskip

The isolation of the superfluous structure hidden behind the
Leibniz equivalence (namely the gauge variables describing
inertial effects) renders even more glaring the ontological
diversity and prominence of the gravitational field with respect
to all other fields, as well as the difficulty of reconciling the
deep nature of the gravitational field with the standard wisdom of
theories based on background space-time like effective quantum
field theory and string theory. Any procedure of linearizing and
quantizing these latter unavoidably leads to looking at gravity as
to a spin-2 theory in which the graviton stands on the same
ontological level of other quanta: in the standard approach,
photons, gluons and gravitons all live on the stage on equal
footing. From the point of view we gained in this paper, however,
quantum DO, i.e. {\it non-perturbative gravitons}, do in fact
constitute the stage for the causal play of photons, gluons as
well as of other matter actors like electrons and quarks. More
precisely, if our main conjecture is sound, the {\it
non-perturbative graviton} would be represented by a pair of {\it
scalar fields}.

\bigskip

B) Let us close this survey with some hints that our results tend
to suggest for the quantum gravity programme. As well-known this
programme is documented nowadays by two inequivalent quantization
methods: i) the perturbative background-dependent either {\it
string} or effective QFT formulations, on a Fock space containing
elementary particles; ii) the non-perturbative
background-independent {\it loop} quantum gravity approach, based
on the non-Fock {\it polymer} Hilbert space. In this connection,
see Ref.\cite{64} for an attempt to define a {\it coarse-grained
structure} as a bridge between standard {\it coherent states} in
Fock space and some {\it shadow states} of the discrete quantum
geometry associated to a {\it polymer} Hilbert space. As
well-known, this approach still fails to accommodate elementary
particles.
\medskip

Now, the individuation procedure we have proposed transfers, as it
were, the {\it non-commutative} Poisson-Dirac structure of the
Dirac observables onto the individuated point-events even if, of
course, the coordinates on the l.h.s. of the identity
Eq.(\ref{IV8}) are c-numbers quantities. Of course, no direct
physical meaning can be attributed to this circumstance at the
classical level. One could guess, however, that such feature might
deserve some attention in view of quantization, for instance by
maintaining that the identity (Eq.(\ref{IV8})) could still play
some role at the quantum level. We will assume here that the {\it
main conjecture} is verified so that all the quantities we
consider are manifestly covariant. On the other hand, this is a
logical necessity in order to get a coordinate-independent
quantization procedure.

\medskip

Let us first lay down some qualitative premises concerning the
status of Minkowski space-time in relativistic quantum field
theory (call it {\it micro space-time}, see Ref.\cite{65}). Such
status is quite peculiar. From the chrono-geometric point of view,
the micro space-time is a universal, classical, non-dynamical
space-time, just Minkowski's space-time of the special theory of
relativity, utilized without any scale limitation from below.
However, it is introduced into the theory through the
group-theoretical requirement of relativistic invariance of the
{\it statistical} results of measurements with respect to the
choice of {\it macroscopic reference frames}. The micro space-time
is therefore {\it anchored} to the macroscopic medium-seized
objects that asymptotically define the experimental conditions in
the laboratory. It is, in fact, in this asymptotic sense that a
physical meaning is attributed to the classical spatiotemporal
{\it coordinates} upon which the quantum fields' operators depend
as {\it parameters}. Thus, the spatiotemporal properties of the
{\it micro Minkowski manifold}, including its basic causal
structure, are, as it were, projected on it {\it from
outside}\footnote{ One should not forget that the Minkowski
structure of the {\it micro-space-time} has been probed down to
the scale of $10^{-18}$ m., yet only from the point of view of
{\it scattering} experiments, involving a limited number of {\it
real} particles.}.

\medskip

Summarizing, the role of {\it Minkowski's micro space-time} seems
to be essentially that of an instrumental {\it external
translator} of the symbolic structure of quantum theory into the
{\it causal} language of the macroscopic irreversible traces that
constitute the experimental findings within {\it macro
space-time}. The conceptual status of this {\it external
translator} fits then very well with that of epistemic
precondition for the formulation of relativistic quantum field
theory in the sense of Bohr, independently of one's attitude
towards the interpretation of quantum theory of measurement. Thus,
barring macroscopic Schr\"odinger Cat states of the would-be
quantum space-time, any conceivable formulation of a quantum
theory of gravity would have to respect, at the {\it operational}
level, the {\it epistemic priority} of a classical spatiotemporal
continuum. Talking about the quantum structure of space-time needs
overcoming a serious conceptual difficulty concerning the
localization of the gravitational field: indeed, what does it even
mean to talk about the {\it values} of the gravitational field
{\it at a point}, to the effect of points individuation, if the
field itself is subject to quantum fluctuations ? One needs in
principle some sort of reference structure in order to give
physical operational meaning to the spatiotemporal language, one
way or the other. It is likely, therefore, that in order to
attribute some meaning to the individuality of points that lend
themselves to the basic structure of standard quantum theory, one
should split, as it were, the individuation of point-events from
the true quantum properties, i.e., from the fluctuations of the
gravitational field and the micro-causal structure.

\medskip

Now, it seems that our canonical analysis of the individuation
issue, tends to prefigure a new approach to quantization having in
view a Fock space formulation. Accordingly, unlike loop quantum
gravity, this approach could even lead to a background-independent
incorporation of the standard model of elementary particles
(provided the Cauchy surfaces admit Fourier transforms). Two
options present themselves for a quantization program respecting
relativistic causality \footnote{ Recall that a 3+1 splitting of
the mathematical space-time, including the notions of space-like,
light-like, and time-like directions, is presupposed from the
beginning.}:
\medskip

\noindent 1) The procedure for the individuation outlined in
Section II suggests to quantize the DO=BO of each Hamiltonian
gauge, as well as all the matter DO, and to use the weak ADM
energy of that gauge as Hamiltonian for the functional
Schr\"odinger equation (of course there might be ordering
problems). This quantization would yield as many Hilbert spaces as
4-coordinate systems, which would likely be grouped in unitary
equivalence classes (we leave aside asking what could be the
meaning of inequivalent classes, if any). In each Hilbert space
the DO=BO quantum operators would be distribution-valued quantum
fields on a {\it mathematical micro space-time} parametrized by
the 4-coordinates $\tau$, $\vec \sigma$ associated to the chosen
gauge. Strictly speaking, due to the non-commutativity of the
operators ${\hat {\tilde F}}{}^{\bar A}$ associated to the
classical gauge-fixing (\ref{IV5}) $\sigma^A - F^{\bar A} \approx
0$ defining that gauge, there would be {\it no space-time manifold
of point-events} to be mathematically identified by one coordinate
chart over the {\it micro-space-time}: only a {\it gauge-dependent
non-commutative structure} which is likely to lack any underlying
topological space structure. However, for each Hilbert space, a
{\it coarse-grained} space-time of point-events might be
associated to each solution of the functional Schr\"odinger
equation, through the expectation values of the operators ${\hat
{\tilde F}}{}^{\bar A}$:

\beq
 {\bar \Sigma}^{\bar A} (\tau, \vec \sigma) = \langle \Psi
\Big| {\tilde F}^{\bar A}_{G}[ {\bf R}^{\bar a}(\tau ,\vec \sigma
), {\bf \Pi}_{\bar a}(\tau , \vec \sigma)] \Big| \Psi
\rangle,\quad a = 1,2,
 \label{VIII1}
  \eeq

\noindent where ${\bf R}^{\bar a}(\tau ,\vec \sigma )$ and ${\bf
\Pi}_{\bar a}(\tau , \vec \sigma)$ are {\it scalar Dirac
operators}.
\medskip

Let us note that, by means of Eq.(\ref{VIII1}), the {\it
non-locality} of the {\it classical} individuation of point-events
would directly get imported at the basis of the ordinary quantum
non-locality.

\medskip
Also, one could evaluate in principle the expectation values of
the operators corresponding to the lapse and shift functions of
that gauge. Since we are considering a quantization of the
3-geometry (like in loop quantum gravity), evaluating the
expectation values of the quantum 3-metric and the quantum lapse
and shift functions could permit to reconstruct a coarse-grained
foliation with coarse-grained WSW hyper-surfaces\footnote{ This
foliation is called \cite{66} the Wigner-Sen-Witten (WSW)
foliation due to its properties at spatial infinity (see footnote
14 of I).}. \bigskip

\noindent 2) In order to avoid inequivalent Hilbert spaces, we
could quantize {\it before} adding any gauge-fixing (i.e.
independently of the choice of the 4-coordinates and the
individuation of point-events), using e.g., the following rule of
quantization, which respects relativistic causality: in a given
canonical basis of the conjecture, quantize the two pairs of DO=BO
observables and the matter DO, but leave the 8 gauge variables
$\zeta^{\alpha}(\tau ,\vec \sigma )$, $\alpha =1,..,8$, as {\it
c-number classical fields} ({\it c-number generalized times}).
Like in Schr\"odinger's theory with time-dependent Hamiltonian,
where $i\, {{\partial}\over {\partial t}}$ is equated to the
action of the Hamiltonian operator, the momenta conjugate to the
gauge variables would be represented by the functional derivatives
$i \delta / \delta \zeta^{\alpha}(\tau ,\vec \sigma )$. Assuming
that, in the chosen canonical basis of our {\it main conjecture},
7 among the eight constraints be gauge momenta, we would get 7
Schr\"odinger equations $i \delta / \delta \zeta^{\alpha}(\tau
,\vec \sigma )\, \Psi( {R}^{\bar a} | \tau; \zeta^{\alpha} ) = 0$
from them. Let ${\sl H}(new) \approx 0$ be the super-Hamitonian
constraint and $E_{ADM}(new)$ the weak ADM energy, in the new
basis. Both would become operators $\hat{\sl H}\,\, or\,\, {\hat
E}_{ADM}({\hat r}_a, {\hat \pi}_a, \zeta^{\alpha}, i \delta /
\delta \zeta^{\alpha})$. If an ordering existed such that the 8
quantum constraints ${\hat \phi}_{\alpha}$ and ${\hat E}_{ADM}$
satisfied a closed algebra $[ {\hat \phi}_{\alpha}, {\hat
\phi}_{\beta}] = {\hat C}_{\alpha\beta\gamma}\, {\hat
\phi}_{\gamma}$ and $[ {\hat E}_{ADM}, {\hat \phi}_{\alpha} ] =
{\hat B}_{\alpha\beta}\, {\hat \phi}_{\beta}$ (with the quantum
structure functions tending to the classical ones for $\hbar
\mapsto 0$), we might quantize by imposing the following 9 coupled
integrable functional Schr\"odinger equations

\bea
 &&i {{\delta} \over
{\delta \zeta_{\alpha}(\tau ,\vec \sigma )}}\, \Psi( {R}^{\bar a}
| \tau; \zeta^{\alpha} ) = 0,\quad \alpha = 1,..,7,\quad
\Rightarrow\,\, \Psi = \Psi ({R}^{\bar a}|\tau ;
\zeta^8),\nonumber \\
 &&{}\nonumber \\
 &&\hat {\sl H}(R^{\bar a}, i {{\delta}\over {\delta R_{\bar a}}}, \zeta^{\alpha}, i
 {{\delta}\over {\delta \zeta^{\alpha}}})\, \Psi ( {R}^{\bar a}
| \tau; \zeta^{\alpha} ) = 0,\qquad \bar a = 1,2,\nonumber \\
 &&{}\nonumber \\
 &&i\, {{\partial}\over {\partial \tau}}\, \Psi ( {R}^{\bar a}
| \tau; \zeta^{\alpha} ) =  {\hat E}_{ADM}(R^{\bar a}, i
{{\delta}\over {\delta R^{\bar a}}}, \zeta^{\alpha}, i
 {{\delta}\over {\delta \zeta^{\alpha}}})\, \Psi( {R}^{\bar a}
| \tau; \zeta^{\alpha} ),
 \label{VIII2}
 \eea

\noindent with the associated usual Schroedinger scalar product $
\langle \Psi \Big| \Psi \rangle$ being independent of $\tau$ and $
\zeta^{\alpha}$'s because of Eq.(\ref{VIII2}). This is similar to
what happens in the quantization of the two-body problem in
relativistic mechanics \cite{62,67,68}.

\medskip

If the previously described quasi-Shanmugadhasan canonical basis
exists, the wave functional would depend on 8 functional field
parameters $\zeta^{\alpha}(\tau ,\vec \sigma )$, besides the
mathematical time $\tau$ (actually only on $\zeta^8$). Each {\it
curve} in this parameter space would be associated to a
Hamiltonian gauge in the following sense: for each solution $\Psi$
of the previous equations, the classical gauge-fixings $\sigma^A -
F_G^{\bar A} \approx 0$ implying $\zeta^{\alpha} =
\zeta^{(G)\alpha}(R^{\bar a}, \Pi_{\bar a})$, would correspond to
expectation values $< \Psi |\, \zeta^{(G)\alpha}(\tau ,\vec \sigma
) | \Psi > =  {\tilde \zeta}^{(G)\alpha}(\tau ,\vec \sigma )$
defining the {\it curve} in the parameter space. Again, we would
have a {\it mathematical micro space-time} and a {\it
coarse-grained space-time of "point-events"}. At this point, by
going to {\it coherent states}, we could try to recover classical
gravitational fields\footnote{ At the classical level, we have the
ADM Poincar\'e group at spatial infinity on the asymptotic
Minkowski hyper-planes orthogonal to the ADM 4-momentum, while the
WSW hyper-surfaces tend to such Minkowski hyper-planes in every
4-region where the 4-curvature is negligible, because their
extrinsic curvature tends to zero in such regions. Thus, matter
and gauge fields could be approximated there by the rest-frame
relativistic fields whose quantization leads to relativistic QFT.
Since at the classical level, in each 4-coordinate system, matter
and gauge field satisfy $\phi (\tau ,\vec \sigma ) = \phi
(\sigma^A) \approx \tilde \phi (F^{\bar A}) = {\tilde {\tilde
\phi}}(R^{\bar a}, \Pi_{\bar a})$, they could be thought of as
functions of either the intrinsic pseudo-coordinates (as DeWitt
does) or the DO=BO observables of that gauge.}. The 3-geometry
(volumes, areas, lengths) would be quantized, perhaps in a way
coherent with the results of loop quantum gravity.

\bigskip

It is important to stress that, according to both of our
suggestions, {\it only the DO would be quantized}. The upshot is
that fluctuations in the gravitational field (better, in the DO)
would entail fluctuations of the point texture that lends itself
to the basic space-time scheme of standard relativistic quantum
field theory: such fluctuating texture, however, could be
recovered as a coarse-grained structure. This would induce
fluctuations in the coarse-grained metric relations, and thereby
in the causal structure, both of which would tend to disappear in
a semi-classical approximation. Such a situation should be
conceptually tolerable, and even philosophically appealing, as
compared with the impossibility of defining a causal structure
within all of the attempts grounded upon quantization of the full
4-geometry.

\medskip

Besides, in space-times with matter, our procedure entails {\it
quantizing the tidal effects and action-at-a-distance potentials
between matter elements but not the inertial aspects of the
gravitational field}. As shown before, the latter are connected
with the gauge variables whose variations reproduce all the
possible viewpoints of local accelerated time-like observers.
Thus, quantizing the gauge variables would be tantamount to
quantizing the metric {\it and} the {\it passive observers} and
their reference frames associated to the congruences studied in
Section IV of I. Of course, such observers have nothing to do with
the {\it dynamical observers}, which should be realized in terms
of the DO of matter.

\medskip

Finally, concerning different ways of looking at inertial forces,
consider for the sake of completeness the few known attempts of
extending non-relativistic quantum mechanics from global inertial
frames to global non-inertial ones \cite{69} by means of
time-dependent unitary transformations $U(t)$. The resulting
quantum potentials $V(t) = i\, \dot U(t)\, U^{-1}(t)$ for the
fictitious forces in the new Hamiltonian $\tilde H = U(t)\, H
U^{-1}(t) + V(t)$ for the transformed Schr\"odinger
equation\footnote{Note that as it happens with the time-dependent
Foldy-Wouthuysen transformation \cite{70}, the operator $\tilde H$
describing the non-inertial time evolution is no more the energy
operator.}, as seen by an accelerated observer (passive view), are
often re-interpreted as action-at-a-distance Newtonian
gravitational potentials in an inertial frame (active view). This
fact, implying in general a change in the emission spectra of
atoms, is justified by invoking an extrapolation of the
non-relativistic limit of the weak equivalence principle
(universality of free fall or identity of inertial and
gravitational masses) to quantum mechanics. Our Hamiltonian
distinction among tidal, inertial and action-at-a-distance effects
supports Synge's criticism \cite{21} b) of Einstein's statements
about the equivalence of {\it uniform} gravitational fields and
{\it uniform} accelerated frames. Genuine physical uniform
gravitational fields do not exist over finite regions\footnote{Nor
is their definition a unambiguous task in general \cite{71}.} and
must be replaced by tidal and action-at-a-distance effects: these,
however, are clearly not equivalent to uniform acceleration
effects. From our point of view, the latter are generated as
inertial effects whose appearance depends upon the gauge
variables. Consequently, the non-relativistic limit of our
quantization procedure should be consistent with the previous
passive view in which atom spectra are not modified by pure
inertial effects, and should match the formulation of standard
non-relativistic quantum mechanics of Ref.\cite{62}.

\vfill\eject

\appendix

\section{Axiomatic Foundations and Theory of Measurement in General Relativity.}

In this Appendix we review the Ehlers-Pirani-Schild axiomatic
approach \cite{26} to the theory of measurement in general
relativity, based on {\it idealized test matter}.

After a critique of the Synge's {\it chronometrical} axiomatic
approach\cite{21} \footnote{Synge accepts as basic primitive
concepts {\it particles} and {\it (standard) clocks}. Then he
introduces the 4-metric as the fundamental structure, postulating
that whenever $x$, $x + dx$ are two nearby events contained in the
world line or history of a clock, then the separation associated
with $(x, x+dx)$ equals the time interval as measured by that (and
by other suitably scaled) clock. These axioms are good for the
{\it deduction} of the subsequent theory, but are not a good {\it
constructive} set of axioms for relativistic space-times
geometries. The Riemannian line element cannot be derived by
clocks alone without the use of light signals. The chronometrical
determination of the 4-metric components does not compellingly
determine the behavior of freely falling particles and light rays
and Synge has to add a further axiom (the geodesic hypothesis). On
the basis of this axiom it is then possible (Marzke \cite{41},
Kundt-Hoffmann \cite{72}) to construct clocks by means of freely
falling particles and light rays (i.e. to give a physical
interpretation of the 4-metric in terms of time). Therefore the
chronometrical axioms appear either as redundant or, if the term
{\it clock} is interpreted as {\it atomic clock}, as a link
between macroscopic gravitation theory and atomic physics: these
authors claim for the equality of gravitational and atomic time.
It should be better to test this equality experimentally (in radar
tracking of planetary orbits atomic time has been used {\it only}
as an {\it ordering parameter}, whose relation to gravitational
time was to be determined from the observations) or to derive it
eventually from a theory that embraces both gravitational and
atomic phenomena, rather than to postulate it as an axiom.},
Ehlers, Pirani and Schild  \cite{26}, reject {\it clocks} as basic
tools for setting up the space-time geometry and propose to use
{\it light rays} and {\it freely falling particles}. The full
space-time geometry can then be synthesized from a few local
assumptions about light propagation and free fall.
\bigskip

a) The propagation of light determines at each point of space-time
the infinitesimal null cone and thus establishes its {\it
conformal structure} ${\cal C}$. In this way one introduces the
notions of being space-like, time-like and null and one can single
out as {\it null geodesics} the null curves contained in a null
hyper-surface (the light rays).

b) The motions of freely falling particles determine a family of
preferred ${\cal C}$-time-like curves. By assuming that this
family satisfies a generalized law of inertia (existence of local
inertial frames in free fall, equality of inertial and passive
gravitational mass), it follows that free fall defines a {\it
projective structure} ${\cal P}$ in space-time such that the world
lines of freely falling particles are the ${\cal C}$-time-like
geodesics of ${\cal P}$.

c) Since, experimentally, an ordinary particle (positive rest
mass), though slower than light, can be made to chase a photon
arbitrarily close, the conformal and projective structures of
space-time are {\it compatible}, in the sense that every ${\cal
C}$-null geodesic is also a ${\cal P}$-geodesic. This makes $M^4$
a {\it Weyl space} $(M^4, {\cal C}, {\cal P})$. A Weyl space
possesses a unique {\it affine structure} ${\cal A}$ such that
${\cal A}$-geodesics coincide with ${\cal P}$-geodesics and ${\cal
C}$-nullity of vectors is preserved under ${\cal A}$-parallel
displacement. In conclusion, light propagation and free fall
define a Weyl structure $(M^4, {\cal C}, {\cal A})$ on space-time
(this is equivalent to an {\it affine connection} due to the
presence of both the projective and the conformal structure).

d) In a Weyl space-time, one can define an {\it arc length}
(unique up to linear transformations) along any non-null curve.
Applying such definition to the time-like world line of a particle
$P$ (not necessarily freely falling), we obtain a {\it proper
time} (= arc length) $t$ on $P$, provided two events on $P$ have
been selected as {\it zero point} and {\it unit point of time}.
The (idealized) Kundt-Hoffmann experiment \cite{72} designed to
measure proper time along a time-like world line in Riemannian
space-time by means of light signals and freely falling particles
can be used without modifications to measure the proper time $t$
in a Weyl space-time.

e) In absence of a {\it second clock effect}\footnote{The first
clock effect is essentially the twin paradox effect. On the other
hand, if the time unit cannot be fixed for all standard clocks
simultaneously in a consistent way, Perlick \cite{34} speaks of a
{\it second clock effect}.} a Weyl space $(M^4, {\cal C}, {\cal
A})$ becomes a Riemannian space, in the sense that there exists a
Riemannian 4-metric ${\cal M}$ compatible with ${\cal C}$ (i.e.
having the same null-cones) and having ${\cal A}$ as its metric
connection. The Riemannian metric is necessarily unique up to a
constant positive factor. Since ${\cal A}$ determines a {\it
curvature tensor} $R$, the use of the equation of geodesic
deviation shows that $(M^4, {\cal C}, {\cal A})$ is Riemannian if
and only if the proper times $t$, $t^{'}$ of two arbitrary,
infinitesimally close, freely falling particles $P$, $P^{'}$ are
{\it linearly} related (to first order in the distance) by {\it
Einstein simultaneity} (see Ref.\cite{26}). In Newtonian
space-time the role of ${\cal C}$ is played by the {\it absolute
time}. It is also easy to add a physically meaningful axiom that
singles out the space-time of special relativity, either by
requiring homogeneity and isotropy of $M^4$ with respect to
$({\cal C}, {\cal A})$, or by postulating vanishing relative
accelerations between arbitrary, neighboring, freely falling
particles.

Now, Perlick \cite{34} states that experimental data on standard
atomic clocks confirm the absence of the {\it second clock
effect}, so that our actual space-time is not Weyl but
pseudo-Riemannian and it is possible to introduce a notion of {\it
ideal rigid rod}.

Let us note that the previous axiomatic approach should be
enlarged to cover tetrad gravity, because of the need of test
gyroscopes to define the triads of the tetrads of time-like
observers. Then the axiomatic would include the possibility of
measuring gravito-magnetism and would have to face the question of
whether or not the free fall of macroscopic test gyroscopes is
geodesic.

An associated theory of the measurement of time-like and
space-like intervals has been developed by Martzke-Wheeler
\cite{33,41}, using Schild geodesic clock (if it is a standard
clock, Perlick's definition of rigid rod can be used): {\it the
axiomatics is replaced by the empirical notion of a fiducial
interval as standard}. Pauri and Vallisneri \cite{73} have further
developed the Martzke-Wheeler approach, showing that, given the
{\it whole} world-line of an accelerated time-like observer, it is
possible to build an associated space-time foliation with
simultaneity space-like non-overlapping 3-surfaces. However,
since, like in the local construction of Fermi coordinate systems,
the 3-surfaces are orthogonal to the observer world-line, its
validity is limited to a neighborhood of the observer, determined
by the acceleration radii. See the discussion in Subsection IIB
and Section VI of Ref.\cite{31}, for the construction of good
foliations with simultaneity 3-surfaces not orthogonal to the
observer world-line.\medskip

As already said, material (test) reference fluids were introduced
by various authors \cite{53,36,59} for simulating the axioms.

\vfill\eject

\end{document}